\documentclass[11pt]{article}
\usepackage{jheppub}
\usepackage[colorinlistoftodos]{todonotes}
\usepackage{tikz}
\usepackage{url}
\usepackage{comment}
\usepackage{color}
\usepackage{xcolor}
\usepackage{soul}
\usepackage{tikz} 
\usepackage{tikz-cd} 
\usepackage{physics}
\usepackage{lscape} 
\usepackage{tabularx}
\usepackage{bm}
\usepackage{mathrsfs}
\usepackage{mathtools}
\usepackage{ulem}

\usepackage{tikz,tikz-3dplot}
\usepackage{pgfplots}
\pgfplotsset{compat=1.17}
\usetikzlibrary{shapes,arrows,cd,chains,decorations.markings,decorations.pathmorphing,calc}
\tikzset{
	->-/.style args={#1rotate#2}{decoration={markings, mark=at position #1 with {\arrow[scale=1.5,rotate = #2 ]{stealth}}}, postaction={decorate}}
}
\tikzset{
	->>-/.style args={#1and#2rotate#3}{decoration={markings, mark=at position #1 with {\arrow[scale=1.5,rotate = #3 ]{stealth}}, mark=at position #2 with {\arrow[scale=1.5,rotate = #3 ]{stealth}}}, postaction={decorate}}
}
\tikzset{
	-s-/.style args={#1rotate#2}{decoration={markings, mark=at position #1 with {\arrow[scale=1.5,rotate = #2]{to}}}, postaction={decorate}}
}
\tikzset{
    marrow/.style={decoration={markings,mark=at position 0.575 with {\arrow{#1}}}, postaction=decorate}
}
\tikzset{
    mmrrow/.style={decoration={markings,mark=at position 0.50 with {\arrow{#1}}}, postaction=decorate}
}
\tikzset{
    farrow/.style={decoration={markings,mark=at position 0.55 with {\arrow[rotate = 180]{#1}}}, postaction=decorate}
}
\tikzstyle{GraphNode}=[circle, draw=black, fill=black, inner sep=2pt, minimum size=5pt]
\tikzstyle{GraphEdge}=[black]
\pgfmathsetmacro{\gS}{1}

\usepackage{tikz} 
\usepackage{tikz-cd} 
\usetikzlibrary{matrix,calc} 
\usetikzlibrary{decorations.markings,shapes.geometric,shapes.misc} 
\usetikzlibrary{decorations.pathreplacing,positioning} 
\usetikzlibrary{arrows}
\usepackage{mathrsfs}

\usepackage{empheq}
\usepackage[most]{tcolorbox}
\newtcbox{\mymath}[1][]{%
    nobeforeafter, math upper, tcbox raise base,
    enhanced, colframe=blue!30!black,
    colback=blue!30, boxrule=1pt,
    #1}

\usepackage{quiver}
\graphicspath{ {./images/} }
\newcommand{\pic}[2]{\vcenter{\hbox{\includegraphics[height=#1]{#2}}}}
\newcommand{\pick}[2]{\vcenter{\hbox{\includegraphics[width=#1]{#2}}}}
\newcommand{\tri}{\pic{1.2ex}{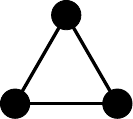}}
\newcommand{\prop}{\pick{1.5ex}{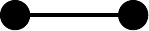}}

\DeclareMathOperator{\calI}{\mathcal{I}}

\DeclareMathOperator{\calN}{\mathcal{N}}
\DeclareMathOperator{\calO}{\mathcal{O}}
\DeclareMathOperator{\calP}{\mathcal{P}}

\DeclareMathOperator{\calS}{\mathcal{S}}
\DeclareMathOperator{\calT}{\mathcal{T}}

\def\bbR{\mathbb{R}}

\def\bbC{\mathbb{C}}

\DeclareMathOperator{\free}{{\mathrm{free}}}

\allowdisplaybreaks

\renewcommand{\arraystretch}{1.5}

\renewcommand{\bm}{\mathcal}

\title{Loop Corrected Supercharges from Holomorphic Anomalies}
\author[{Q_0}]{Kasia Budzik}
\author[{Q_1}]{and Justin Kulp}

\affiliation[{Q_0}]{Center for the Fundamental Laws of Nature, Harvard University, Cambridge, MA 02138, USA}
\affiliation[{Q_1}]{Simons Center for Geometry and Physics, 
Stony Brook University, Stony Brook, NY 11794, USA\\
Yang Institute for Theoretical Physics, 
Stony Brook University, Stony Brook, NY 11794, USA}

\emailAdd{kbudzik@fas.harvard.edu}
\emailAdd{jkulp@scgp.stonybrook.edu}

\abstract{
We describe the loop corrections to supercharges in supersymmetric quantum field theories using the holomorphic twist formalism. We begin by reviewing the relation between supercharge corrections and the ``twice-generalized'' Konishi anomaly, which corrects the \textit{semi}-chiral ring. In the holomorphic twist, these corrections appear as BRST anomalies and are computed using the higher operations of an underlying $L_\infty$ conformal algebra. We then apply this formalism to obtain the complete one-loop corrections to the supercharge of four-dimensional Lagrangian supersymmetric gauge theories, including $\mathcal{N}=4$ SYM, where it admits a remarkably compact expression in terms of superfields.

}

\begin{document}

\maketitle

\section{Introduction and Summary}

The tractability of Supersymmetric Quantum Field Theories (SQFTs) comes from the existence of $\bm{Q}$-closed local operators, i.e. those annihilated by the action of some (set of) supercharge(s) 
\begin{equation}
    \bm{Q}\! \calO = 0\,.
\end{equation}
Such operators are useful for understanding SUSY vacua \cite{Seiberg:1994bz, CDSW}, deformations and non-renormalization theorems \cite{Sohnius:1981sn, deWit:1984rvr, Seiberg:1988ur, Seiberg:1993vc}, and for matching across supersymmetric and holographic dualities \cite{Kutasov:1995ve, Kutasov:1995np, Kutasov:1995ss, Lee:1998bxa, Aharony:1999ti, Romelsberger:2007ec, Dolan:2008qi}. Thus, to understand the space of $\bm{Q}$-closed operators is to understand the ``superdata'' of an SQFT.

In this paper, we consider the most general set of such operators in (3+1)d SQFTs: the semi-chiral operators. By definition, semi-chiral operators are annihilated by one (minimal rank) nilpotent supercharge, say\footnote{The precise choice of supercharge is immaterial \cite{Eager:2018dsx, Elliott:2020ecf, Budzik:2023xbr}, so long as the rank is minimal. Our formalism extends straightforwardly to other dimensions and twists, as described in \cite{Gaiotto:2024gii}. However, the main focus of this note concerns loop corrections to $\bm{Q}$ in (3+1)d gauge theories.}
\begin{equation}
    \bm{Q} := Q_{-}^1\,.
\end{equation}
Correlation functions of semi-chiral operators are regular and insensitive to changes by $\bm{Q}$-exact quantities in SUSY vacua. Consequently, it is useful to pass to $\bm{Q}$-cohomology, turning our attention to the ``semi-chiral ring'' of the SQFT -- a holomorphic generalization of the chiral ring.

In this note, we describe a systematic procedure for computing the quantum corrected semi-chiral ring of interacting SQFTs using a perturbative expansion:
\begin{equation}
    \bm{Q} = Q_{\text{free}} + Q_{\text{tree}} + \hbar Q_{1\text{-loop}} + \hbar^2 Q_{2\text{-loop}} + \dots \,, \label{eq:Qexpintro}
\end{equation}
based on the holomorphic twist formalism of \cite{Budzik:2022mpd, Budzik:2023xbr, Gaiotto:2024gii,Bomans:2025klo}.

Before discussing SQFTs, it is instructive to consider an analog of these ideas in Conformal Field Theory (CFT). In CFTs, the scaling dimensions $\Delta_i$ and three-point functions $C_{ijk}$ of primary local operators make up the entirety of the conformal data. By the state-operator correspondence, understanding the space of local operators also tells us about all states in the (vacuum sector of the) QFT, so enumerating the space of local operators is as complicated as computing the partition function of the CFT on $S^1 \times S^d$.

Quantum corrections to the action of the dilatation operator $D$ on local operators are described by anomalies/anomalous dimensions. In perturbation theory, we construct composite local operators from normal ordered product of free fields, i.e. given fields $\phi_1$ and $\phi_2$, we can construct the normal-ordered primary $:\!\!\phi_1\phi_2\!\!:$, and so on with derivatives. The free dilatation operator $D_{\free}$ acts on normal-ordered products by the Leibniz rule, so $:\!\!\phi_1\phi_2\!\!:$ has classical free scaling dimension $\Delta_1 + \Delta_2$. However, in the interacting quantum theory we can have multiple Wick contractions, leading to loop diagrams and thus anomalous dimensions
\begin{equation}
    D\!:\!\!\phi_1\phi_2\!\!: \,= (\Delta_1 + \Delta_2 + \gamma_{12}(\hbar))\!:\!\!\phi_1\phi_2\!\!: \,,
\end{equation}
i.e., operator renormalization generically prevents $D$ from acting by the Leibniz rule. 

The connection between modified actions of generators, anomalies, and quantum corrected equations of motion is neither exotic nor limited to $D$. Even the action of translations on operators can receive quantum corrections. For example, any symmetry current $J^\mu$ is classically conserved $P_\mu J^\mu = 0$ using the equations of motion, but can be non-conserved quantum mechanically. For chiral fermions in a representation $R$, the action of $P_\mu$ on the current $J_5^\mu$ famously receives corrections at one-loop
\begin{equation}\label{eq:chiralAnomaly}
    P_\mu J_{5}^\mu = g^2 \hbar \frac{2C(R)}{16\pi^2} \tr F_{\mu\nu} \tilde{F}^{\mu\nu}\,.
\end{equation}

As with $D$ and $P_\mu$, the quantum corrections to the action of $\bm{Q}$ in \eqref{eq:Qexpintro} are understood as ``Konishi anomalies.'' \cite{Konishi:1983hf, Konishi:1985tu}. Indeed, it has long been known that the ``generalized Konishi anomaly'' modifies identities in the chiral ring of $\calN=1$ theories \cite{CDSW}, giving corrections to $\bm{Q}$ -- briefly reviewed in Section \ref{sec:QAsAnomaly}. In \cite{Budzik:2023xbr} and in this paper, we compute modifications to identities in the semi-chiral ring, thereby giving a ``twice-generalized Konishi anomaly.''

A general prescription for computing the perturbative corrections to the supercharge was proposed in \cite{Budzik:2022mpd, Budzik:2023xbr} using the holomorphic twist formalism (further developed in \cite{Gaiotto:2024gii, Bomans:2025klo, Gui:2025dqp}). In this approach, one adds $\bm{Q}$ to the BRST charge of a supersymmetric gauge theory, producing a simplified theory with holomorphic dependence on spacetime coordinates. For Lagrangian SQFTs, the twisted theory can be further mapped to vastly simpler holomorphic $\beta\gamma$-$bc$ systems \cite{Costello:2013zra, Saberi:2019ghy, Elliott:2020ecf}. The twisted theories also possess an $L_\infty$ conformal algebra structure, generated by algebraic operations called higher $\lambda$-brackets, and can admit infinite-dimensional symmetries generalizing Virasoro and Kac-Moody symmetries to higher dimensions \cite{faonte2019higher, Gwilliam:2018lpo, Saberi:2019fkq, Bomans:2023mkd, Budzik:2023xbr, Williams:2024mcc, Scheinpflug:2024mtn, Chen:2025ujx, Bomans:2025klo}.

The aforementioned $\lambda$-brackets provide a controlled perturbative expansion for the holomorphic BRST anomalies of the free theory under deformations. In particular, the higher $\lambda$-brackets provide an explicit formula for the $n$-loop corrections to the supercharge
\begin{align}
Q_n \mathcal{O} = \frac{1}{(n+1)!} \{ \underbrace{\mathcal{I} , \dots , \mathcal{I}}_{n+1} , \mathcal{O} \} \, , \label{eq:Qnintro}
\end{align}
where $\calI$ is an interaction term and the precise definition of the higher bracket is given in Section \ref{sec:QAsAnomaly} \cite{Budzik:2022mpd, Budzik:2023xbr, Gaiotto:2024gii, Gui:2025dqp}. As we explain later, the corrections $Q_n$ can act on $n+1$ letters, and therefore the quantum corrected supercharge does not satisfy the Leibniz rule on local operators in this scheme.

The higher-loop bracket \eqref{eq:Qnintro} and the twice-generalized Konishi anomaly were already employed in \cite{Budzik:2023xbr} in the context of pure $\mathcal{N}=1$ SYM with gauge group $SU(N)$. There it was observed that the tree-level $Q$-cohomology receives quantum corrections both in the planar limit and at finite $N$. In the planar limit, the tree-level cohomology can be computed using homological algebra methods (the Loday-Quillen-Tsygan theorem \cite{Loday1984CyclicHA, Tsygan1983TheHO}), and is generated by three infinite towers of operators:\footnote{As reviewed in Section \ref{sec:Lagrangians}, $b$ can be identified with the holomorphic field strength $F_{++}$ and $\partial_{\dot\alpha}c$ with the gauginos $\bar\lambda_{\dot\alpha}$.}
\begin{equation}
    A_n = \Tr b^n\,,\quad
    B_{n,\dot\alpha} = \Tr b^n \partial_{\dot\alpha} c\,,\quad
    C_n = \Tr b^n \partial_{\dot\alpha}c \partial^{\dot\alpha}c\,,
\end{equation}
and their holomorphic derivatives $\partial_{z^{\dot\alpha}}$. At one-loop, $Q_1$ pairs these towers non-trivially, with 
\begin{equation}
    Q_1 A_n \sim \partial^{\dot\alpha} B_{n,\dot\alpha} 
    \quad\text{and}\quad
    Q_1 B_{n,\dot\alpha} \sim \partial_{\dot\alpha} C_n \,.
\end{equation}
Consequently, the $A_n$ and $B_{n,\dot\alpha}$ towers are no longer closed and the derivative towers $\partial^{\dot\alpha} B_{n,\dot\alpha}$ and $\partial_{\dot\alpha}C_n$ become $Q_1$-exact. The paired towers are lifted from the classical cohomology, leaving only the $C_n$ tower (without derivatives) in the quantum-corrected cohomology.\footnote{At $N=2$, there are additional operators in the $Q_1$-cohomology besides the $C_n$ tower, indicating that the phenomenon of fortuity (discussed later) may also arise after loop corrections. It was further conjectured that the ``one-loop fortuitous" operators in this example are lifted by a non-perturbative differential $Q_{\text{np}}$ \cite{Budzik:2023xbr}.} In this example, the twice-generalized Konishi anomaly manifests as the failure of the stress tensor $B_{1,\dot\alpha}=\Tr b\partial_{\dot\alpha}c$ to be closed under the 1-loop supercharge $Q_1$ \cite{Budzik:2023xbr}.

The study of $\bm{Q}$-cohomology has recently experienced a resurgence in interest due to its relevance in the construction of BPS black hole microstates. In SCFTs, the $\bm{Q}$-cohomology classes are in one-to-one correspondence with $1/4\calN$-BPS states.\footnote{$1/4\mathcal{N}$-BPS operators are by definition annihilated by one $Q$ and $S=Q^\dagger$. In each $Q$-cohomology class there is a unique representative that is also annihilated by $S$.} In $\mathcal{N}=4$ SYM at finite $N$, two classes of $1/16$-BPS operators were recently identified: ``monotone'' operators, which are BPS at all values of $N$; and ``fortuitous'' operators, which become BPS only at special finite values of $N$ due to trace relations \cite{Chang:2022mjp, Choi:2022caq, Choi:2023znd, Budzik:2023vtr, Choi:2023vdm, Chang:2024zqi, deMelloKoch:2024pcs, Gaikwad:2025ugk} (see also \cite{Chang:2024lxt, Chen:2025sum, Chang:2025rqy, Kim:2025vup, Chang:2025wgo, belin2025fortuityabjm} for other examples). Recently, it was observed that the classical $1/16$-BPS spectrum of $\mathcal{N}=4$ SYM with $SO(7)$ gauge group also receives loop corrections \cite{Chang:2025mqp, Choi:2025bhi}, violating non-renormalization conjectures \cite{Grant:2008sk, Chang:2022mjp}. The one-loop corrections have been found to lift two operators from the tree-level $Q$-cohomology \cite{Choi:2025bhi}: one fortuitous (identified in \cite{Gadde:2025yoa, Chang:2025mqp}) and one monotone (identified in \cite{Chang:2025mqp}).

In this note, using the holomorphic twist formalism, we compute the full one-loop corrections to the supercharge in four-dimensional Lagrangian theories whose field content consists of a vector multiplet, an arbitrary number of chiral multiplets, and a superpotential $W$. In particular, we apply our results to $\mathcal{N}=4$ SYM and present the complete one-loop supercharge (Section \ref{sec:Q1computation} and Appendix \ref{app:derivatives}), one term of which was computed in \cite{Choi:2025bhi}.

In the case of $\mathcal{N}=4$ SYM, we find an interesting reorganization of the one-loop result. The $1/16$-BPS fields can be combined into a superfield $C(z,\theta)$ on the superspace $\mathbb{C}^{2|3}$ and the classical action of the supercharge can be written as \cite{Chang:2013fba}
\begin{align}
    Q_0 C(\theta) = \frac{1}{2}[C(\theta),C(\theta)] \, . \label{eq:Q0C}
\end{align}
Remarkably, we observe that the one-loop corrections can also be reorganized into the compact form\footnote{The full action of $Q_1$ includes also the action on fields with derivatives, as recorded in Appendix \ref{app:derivatives}.}
\begin{equation}
    Q_1 (C^A(\theta) C^B(\theta')) = -\frac{1}{2} f_{ACD}f_{BCE}  (\theta_1-\theta_1')(\theta_2-\theta_2')(\theta_3-\theta_3') \partial_{\dot\alpha} C^D(\theta) \partial^{\dot\alpha} C^E(\theta')\,. \label{eq:Q1C}
\end{equation} 
A similar simplification also occurs for $\mathcal{N}=1$ and $\mathcal{N}=2$ pure SYM, see Sections \ref{sec:N1} and \ref{sec:N2}.\footnote{The one-loop correction to the BRST operator in the chiral algebra subsector of $\mathcal{N}=4$ SYM \cite{Beem:2013sza} was computed in \cite{Chang:2023ywj} and also admits a similar repackaging
    \begin{align}
        Q_1^{2d} (C^A(\theta)C^B(\theta')) = -2 f_{ABC} (\theta_1-\theta_1')(\theta_2-\theta_2') \partial C^C(\theta) \,.
    \end{align}}

The operator \eqref{eq:Q0C} is the standard Lie algebra cohomology differential. It would be interesting to understand the mathematical structure corresponding to the deformation \eqref{eq:Q1C}.

It is an open problem whether the loop corrections to the supercharge computed in this note modify the classical BPS spectrum of $\mathcal{N}=4$ SYM with gauge group $SU(N)$; equivalently, whether the $SU(N)$ $Q_1$-cohomology is isomorphic to the $Q_0$-cohomology. We present an argument that the one-loop corrections do not modify the infinite-$N$ single-trace cohomology. At finite $N$, it is natural to ask whether the fortuitous operators can be lifted. One can rule out by computer search \cite{Chang:2025mqp} and by directly computing the one-loop action \cite{Budzik:2023vtr} that the lowest $SU(2)$ fortuitous cohomology \cite{Chang:2022mjp, Choi:2022caq} survives.

In the approach developed in \cite{Budzik:2022mpd, Budzik:2023xbr} and studied in this note, higher loop corrections \eqref{eq:Qnintro} are necessarily present, which can be seen e.g. by checking that $Q_1$ does not square to zero\footnote{Note that $Q_1$ does square to zero \textit{on the $Q_0$-cohomology}, so the prescription of taking $Q_1$-cohomology on the $Q_0$-cohomology is well defined.}
\begin{align}
    Q_1^2 + \qty{Q_0, Q_2} = 0 \, ,
\end{align}
and thus a non-trivial $Q_2$ has to be present.\footnote{However, one can prove by graph combinatorics that for an operator of length $n$ the possible corrections truncate at at most $Q_{n+1}$, but based on computing explicit integrals we expect them to truncate at $Q_{n-1}$.} The higher $\lambda$-brackets, and hence the operators $Q_n$, depend on the choice of scheme. Nevertheless, the lifting of cohomology classes from the final perturbative cohomology (as in the examples \cite{Budzik:2023xbr, Choi:2025bhi}) is scheme independent.\footnote{Under the assumption that non-perturbative corrections to the supercharge may only further reduce the cohomology.} It remains an interesting problem whether the higher loop corrections to the supercharge modify the BPS spectrum and whether they could be
removed using a different (theory-specific) scheme.

\subsubsection*{Outline} 
The outline of the rest of the paper is as follows.
\begin{enumerate}
    \item[{\hyperref[sec:QuantumCorrections]{\underline{$\S$.2.}}}] In Section \ref{sec:QuantumCorrections} we review the algebraic structures we will use in the remainder of the paper. In Section \ref{sec:SemiChiralOps} we start by defining the semi-chiral ring and the holomorphic twist. Then, in Section \ref{sec:QAsAnomaly}, we review how quantum corrections to $\bm{Q}$ can be understood as a BRST anomaly given by the $L_\infty$-algebra brackets
    \begin{equation}
        \bm{Q}\! \calO =  \{\calI, \calO\}_0 + \frac{1}{2!}\{\calI,\calI,\calO\}_0 + \dots\,.
    \end{equation}
    Finally, in Section \ref{sec:ComputingLoopCorrections} we explain an explicit scheme to compute the brackets, and thus the corrections to $\bm{Q}$ in holomorphic QFTs, by evaluating a master integral $\calI_{\tri}[\lambda;z]$.
    \vskip 0.25cm
    
    \item[{\hyperref[sec:Lagrangians]{\underline{$\S$.3.}}}] In Section \ref{sec:Lagrangians} we consider the holomorphic twist of Lagrangian SUSY gauge theories with arbitrary gauge group vector multiplet, and any number of chiral multiplets and superpotential. In Section \ref{sec:Q1computation} we use our scheme to obtain the explicit one-loop supercharge $Q_1$ acting on products of fields and their derivatives. We then apply this to the examples of $\calN=1$ (Section \ref{sec:N1}), $\calN=2$ (Section \ref{sec:N2}), and $\calN=4$ (Section \ref{sec:N4}) Super Yang-Mills theories respectively and show the repackaging of the one-loop supercharge as a local expression on superspace.
    \vskip 0.25cm
    
    \item[{\hyperref[app:integral]{\underline{$\mathscr{A}$.}}}] In Appendix \ref{app:integral} we give the explicit computation of the universal 4d holomorphic (theory-independent) one-loop anomaly integral that we employ in the rest of the paper.

    \item[{\hyperref[app:derivatives]{\underline{$\mathscr{B}$.}}}] In Appendix \ref{app:derivatives} we record the explicit action of the one-loop supercharge on operators built from fields with derivatives. We show the derivation in Appendix \ref{app:CombinatorialIdentity}.

\end{enumerate}

\section{The Holomorphic Twist and Corrections to \texorpdfstring{$\bm{Q}$}{Q} from Anomalies}\label{sec:QuantumCorrections}
In this section we define the basic objects that we will use in the next section of the paper. In Section \ref{sec:SemiChiralOps}, we introduce semi-chiral superfields and the holomorphic twist used for computations in the rest of the paper. In Section \ref{sec:QAsAnomaly}, we explain how corrections to the action of $\bm{Q}$ on local operators should be understood as a ``twice generalized Konishi anomaly,'' by reviewing the famous Konishi anomaly corrections to the chiral ring \cite{CDSW} and then the extension to the semi-chiral ring via the holomorphic twist developed in \cite{Budzik:2022mpd, Budzik:2023xbr, Gaiotto:2024gii}. After reviewing this background and justifying the relationship to anomalies, we explain the extremely constrained form of loop corrections in cohomological holomorphic(-topological) QFTs, and thus to the action of $\bm{Q}$ on operators, in Section \ref{sec:ComputingLoopCorrections}.

\subsection{Semi-Chiral Superfields and \texorpdfstring{$\bm{Q}$}{Q}-Cohomology}\label{sec:SemiChiralOps}
We consider $\calN$-extended supersymmetric theories in (3+1)d Euclidean spacetime. We define our supercharges so that
\begin{align}
    \{Q^i_\alpha, \bar{Q}^j_{\dot\beta}\} 
        &= \delta^{ij}P_{\alpha\dot\beta}\,,\\
    \{Q^i_\alpha, Q^j_{\beta}\} 
        &= \{\bar{Q}^i_{\dot\alpha}, \bar{Q}^j_{\dot\beta}\} = 0\,,
\end{align}
where $i,j=1,\dots, \calN$, the $P_{\alpha\dot\beta}$ are translations, and we have used the two-component spinor notation.\footnote{There are also R-symmetries depending on $\calN$ and possible superconformal extensions. However, since we are only interested in computing $Q$-cohomology, the existence and details of twisting homomorphisms and/or rigid SUGRA backgrounds is immaterial.} Without loss of generality \cite{Eager:2018dsx, Elliott:2020ecf}, we define our supercharge to be:
\begin{equation}
    \bm{Q} := Q_{-}^{1}\,.
\end{equation}
This choice explicitly breaks the left $SU(2)_-$ symmetry of $\mathrm{Spin}(4) \cong SU(2)_- \times SU(2)_+$ and distinguishes holomorphic and anti-holomorphic coordinates on spacetime $\bbR^4 \cong \bbC^2$ by
\begin{equation}
    z^{\dot\alpha} := x^{+\dot\alpha}\,,\quad
    \bar{z}^{\dot\alpha} := x^{-\dot\alpha}\,.
\end{equation}
We note that the anti-holomorphic translations $P_{-\dot\alpha}$ are $\bm{Q}$-exact, in the sense that
\begin{equation}\label{eq:QHolomorphic}
    \{\bm{Q}, \bar{Q}_{\dot\alpha}^{1}\} = P_{-\dot{\alpha}}\,.
\end{equation}

The \textit{semi-chiral operators} are defined as the local operators $\calO^{(0)}$ annihilated by $\bm{Q}$, i.e.
\begin{equation}
    [\bm{Q}, \calO^{(0)}] = 0\,.
\end{equation}
In a SUSY vacuum, any $\bm{Q}$-exact term vanishes in correlation functions, implying that correlation functions only depend on the $\bm{Q}$-cohomology class of local operators. The collection of $\bm{Q}$-closed semi-chiral operators modulo $\bm{Q}$-exact semi-chiral operators defines the \textit{semi-chiral ring}. We define the ``holomorphic twist'' of an SQFT to be the cohomological holomorphic QFT obtained by passing to $\bm{Q}$-cohomology in the original SQFT.

Naively, by \eqref{eq:QHolomorphic}, we might have expected the correlation functions of semi-chiral operators to be holomorphic and diverge as local operators become close in $z_{ij}^{\dot\alpha} \to 0$. However, Hartogs' theorem implies that holomorphic correlation functions on $\bbC^{d>1}$ are actually free from singularities. As a corollary, this theorem then implies that the singular part of the OPE of $\bm{Q}$-closed operators in a (3+1)d SQFT must actually be $\bm{Q}$-exact, i.e. vanishing in cohomology. In order to have interesting correlation functions, and compute corrections to $\bm{Q}$, it is necessary to consider correlation functions of $\bm{Q}$-closed operators with supersymmetric descendants of $\calO^{(0)}$. In the cohomological theory, this is captured by cohomological descent.

To understand this, it is useful to introduce a superspace formalism for the $\calN=1$ superalgebra generated by the $Q_\alpha^1$ and $\bar{Q}_{\dot\alpha}^1$, as they are distinguished by our choice of $\bm{Q}$. In particular, given \textit{any} local operator $\calO^{(0)}$, i.e. semi-chiral or not, we can define a reduced $\mathcal{N}=1$ superfield
\begin{equation}
    \calO := e^{i \dd\bar{z}^{\dot\alpha} \bar{Q}^1_{\dot\alpha}} \calO^{(0)} =: \calO^{(0)} + \calO^{(1)} + \calO^{(2)}\,.
\end{equation}
This superfield is a generating function for the $\bar{Q}_{\dot\alpha}^1$-descendants of $\calO^{(0)}$. Instead of introducing Grassmann variables, we have used anti-holomorphic forms $\dd \bar{z}^{\dot\alpha}$ on $\bbC^2$ to organize the superfield components.\footnote{Traditionally, this is done by introducing Grassmann odd coordinates $(\theta^\alpha, \bar{\theta}^{\dot\alpha})$ on $\bbR^4$, making $\bbR^4 \rightsquigarrow \bbR^{4|4}$ so that the $Q_{\alpha}^1$ and $\bar{Q}_{\dot\alpha}^1$ act by Grassmann derivatives on spacetime. Instead, we have noted that $\bar{\theta}^{\dot\alpha}$ transforms like an anti-holomorphic one-form and so replaced $\bar{\theta}^{\dot\alpha} \rightsquigarrow i\dd \bar{z}^{\dot\alpha}$ on $\bbC^{2}$. See e.g. \cite{Axelrod:1991vq, Budzik:2023xbr, Bomans:2025klo}.} In other words, the superfield is a Dolbeault form $\calO \in \Omega^{0,\bullet}(\bbC^2)$, and the superscript denotes an operator's anti-holomorphic form degree or depth in the $\calN=1$ SUSY multiplet. 

Now we define the Dolbeault differential and supercovariant derivative respectively by:
\begin{equation}
    \bar\partial = \dd \bar{z}^{\dot\alpha} \partial_{\bar{z}^{\dot\alpha}}
    \quad\text{and}\quad
    D_- = \bm{Q} + \bar\partial\,.
\end{equation}
A straightforward calculation shows that the superfield components satisfy descent relations
\begin{equation}\label{eq:DescentRelation}
    \bm{Q}\!\calO^{(k)}+\,\bar\partial\!\calO^{(k-1)} = (\bm{Q}\!\calO)^{(k)}\,.
\end{equation}
A special role is played by generating functions/superfields built from semi-chiral operators: a superfield $\calO$ whose bottom component $\calO^{(0)}$ is semi-chiral satisfies
\begin{equation}
    \bm{Q}\!\calO^{(k)} = -\bar\partial\! \calO^{(k-1)}\,,
\end{equation}
i.e. $D_-\!\calO = (\bm{Q}+\bar\partial)\!\calO = 0$; we call these \textit{semi-chiral superfields}. Cohomology classes of $\bm{Q}$ are in one-to-one correspondence with semi-chiral superfields modulo $(\bm{Q}+\bar\partial)$-exact superfields.

Returning to our original point, the descent relations highlight the importance of the cohomological nature of the holomorphic twist: even though the singular part of the OPE of $\bm{Q}$-closed operators $\calO^{(0)}$ is $\bm{Q}$-exact (and thus trivial in $\bm{Q}$-cohomology), correlation functions involving descendants $\calO^{(k)}$ of $\bm{Q}$-closed operators are non-trivial. In the holomorphic twist, these are described by the higher-form components of superfields on $\bbC^2$, and the usual OPE is generalized by integrating the higher-form fields around surfaces. It is therefore important to understand operations on these form-valued fields to extract perturbative corrections to $\bm{Q}$, as we will now explain.

\subsection{Corrections to \texorpdfstring{$\bm{Q}$}{Q} from Anomalies}\label{sec:QAsAnomaly}
In the previous section, we explained how semi-chiral operators and their descendants are described by a cohomological holomorphic QFT, obtained as the $\bm{Q}$-cohomology of the original theory. Here, we will explain how corrections to $\bm{Q}$ can be understood as anomalies in SQFTs, and how these anomalies are neatly captured (and very rigidly constrained) by the ``higher operations'' of the holomorphic twist, see also \cite{williams2020renormalization, Budzik:2023xbr, Gaiotto:2024gii, Bomans:2025klo, Gui:2025dqp}.

\subsubsection{Warmup: The Generalized Konishi Anomaly and the Chiral Ring}
One of the most well-known examples of quantum corrections to $\bm{Q}$ comes from considering the chiral ring in supersymmetric gauge theories, and the modification of equivalence relations by the Konishi anomaly \cite{Konishi:1983hf, Konishi:1985tu, CDSW}. 

By definition, a \textit{chiral operator} $\calO^{(0)}$ is annihilated by all $\calN=1$ supercharges of one chirality
\begin{equation}
    [Q_{\alpha}^1,\calO^{(0)}] = 0\,.
\end{equation}
This is a specialization of our previously defined semi-chiral operator. In a SUSY vacuum, correlation functions of chiral operators are position independent
\begin{equation}
    \expval*{\calO_1^{(0)}(x_1) \cdots \calO_n^{(0)}(x_n)} = \expval*{\calO_1^{(0)} \cdots \calO_n^{(0)}}\,,
\end{equation}
and, in a clustering vacuum, decompose into products of one-point functions \cite{Novikov:1983ee, CDSW, Hori:2003ic}. Thus, we usually only consider operators in the chiral ring modulo $Q_{\alpha}^1$-exact terms, which vanish in SUSY vacua.\footnote{It is actually somewhat subtle to correctly characterize these operators in a cohomological way, unlike the holomorphic twist. One putative definition is that chiral ring consists of all operators in the semi-chiral ring which are $SU(2)_-$ singlets \cite{Budzik:2023xbr}.} Physically, the $Q_{\alpha}^1$-closed chiral operators modulo $Q_{\alpha}^1$-exact chiral operators have the same vev in the same SUSY vacuum, and so characterize the vevs/vacuum structure of the supersymmetric theory, called the \textit{chiral ring}.

Classically, the chiral ring is generated by (gauge invariant) words in the chiral superfields. Quantum mechanically, the action of the $Q^i_{\alpha}$ receive perturbative and non-perturbative corrections e.g. due to loop and instanton effects, (generically) thinning out the list of chiral operators, and modifying the vacuum structure of the classical theory. For example, in pure SYM, the classical chiral ring is generated by the single-trace glueball superfields
\begin{equation}
    C = \Tr \overline{\mathcal{W}}_{\dot\alpha} \overline{\mathcal{W}}^{\dot\alpha}\,,
\end{equation}
subject to the relation $C^h = 0$, where $h$ is the dual Coxeter number of $G$ \cite{CDSW, Witten:2003ye, Etingof:2003dd, Cederwall:2023lev}. In pure SYM, quantum corrections are perturbatively absent but instanton corrections are known to give a vev $\expval*{C^h} = c \Lambda^{3h}$ to the glueballs \cite{Novikov:1983ee, CDSW} (see VII.2.2 of \cite{Dorey:2002ik} for the exact prefactor).

In theories with matter fields, the story changes in a number of interesting ways. First, and most obviously, we have a larger space of free fields to build gauge-invariant chiral words from, see e.g. \cite{CDSW, susyGTMM, Seiberg:2002jq} for a list of generators and relations between them. Second, non-trivial superpotentials modify classical equations of motion, giving interactions which correct the free supercharge(s). And third, the equations of motion are modified quantum mechanically by the Konishi anomaly. Of course, there can be non-perturbative corrections as well, which we anticipate from matching of dualities (see e.g. \cite{Tachikawa:2018sae} and references within).

Let us unpack how these corrections to $Q_{\alpha}^1$ are connected to the Konishi anomaly. In its most fundamental form, a Konishi anomaly is any supersymmetrized version of a one-loop triangle diagram, where (schematically)
\begin{equation}\label{eq:veryKonishi}
    \Tr F_{\mu\nu} \tilde{F}^{\mu\nu} \rightsquigarrow \Tr \overline{\mathcal{W}}_{\dot\alpha} \overline{\mathcal{W}}^{\dot\alpha}\,.
\end{equation}
For example, the usual chiral anomaly \eqref{eq:chiralAnomaly} is supersymmetrized in $\calN=1$ $SU(N)$ SQCD with zero superpotential to
\begin{equation}\label{eq:KonishiAnomaly}
    D^2\bar{\mathcal{J}} = g^2\hbar \frac{2C(R)}{16\pi^2} \Tr \overline{\mathcal{W}}_{\dot\alpha} \overline{\mathcal{W}}^{\dot\alpha}\,,
\end{equation}
linking the glueball superfield $C$ to the (non-)conservation of the current $\bar{\mathcal{J}} = \Tr \Phi e^V \bar\Phi$, which generates rescalings $\delta \bar{\Phi} = \epsilon \bar{\Phi}$ of chiral fields. 

This has important consequences for the chiral ring. Since the left-hand side of \eqref{eq:KonishiAnomaly} is $Q_\alpha^{1}$-exact, the Konishi anomaly implies $C\sim 0$ at one-loop in the chiral ring of SQCD with zero superpotential. More generally, one can add a non-trivial superpotential $W$ and consider even more general holomorphic variations in field space, e.g. $\delta_n \bar{\Phi} = \epsilon \bar{\Phi}^n$, modifying \eqref{eq:KonishiAnomaly} to \cite{CDSW, Tachikawa:2018sae}:
\begin{equation}
    D^2\bar{\mathcal{J}}_n = \bar{\Phi}^n \frac{\partial W}{\partial \Phi} + g^2\hbar \frac{n}{16\pi^2} \Tr \overline{\mathcal{W}}^{\dot\alpha} \overline{\mathcal{W}}_{\dot\alpha} \bar{\Phi}^{n-1}\,.
\end{equation}
This is interpreted as both classical and one-loop quantum corrections to the action of supercharges, sometimes called the ``generalized Konishi anomaly.''

Let us note one important example of the Konishi anomaly changing the \textit{semi}-chiral ring in this classical language.\footnote{As another example, in \cite{Choi:2025bhi}, the Konishi anomaly was shown to lift a fortuitous cohomology in the semi-chiral ring and a monotone cohomology in the chiral ring of $\calN=4$ SYM with gauge group $SO(7)$.} At minimum, any local SQFT admits an $\calS$-supermultiplet containing the SUSY current $S_{\mu\alpha}$ and stress tensor $T_{\mu\nu}$ \cite{Dumitrescu:2011iu}. In particular, we define
\begin{equation}
    S^{(0)}_{\dot\alpha} := S_{+\dot\alpha,+}\,,
\end{equation}
whose $\calN=1$ chiral one-form component $S^{(1)}_{\dot\alpha}$ includes the holomorphic part of the physical stress tensor (see e.g. \cite{Budzik:2023xbr, Bomans:2023mkd, Bomans:2025klo} or \cite{Closset:2013vra, Closset:2014uda, Dumitrescu:2016ltq} for classical SUSY language), and generates the surviving rotations, translations, etc. of $\bbC^2$ in the holomorphic twist. In theories with a $U(1)_R$ symmetry, this field is semi-chiral and thus is in the $\bm{Q}$-cohomology. Without a $U(1)_R$ symmetry, only $\partial^{\dot\alpha} S_{\dot\alpha}$ is semi-chiral and in $\bm{Q}$-cohomology. 

As a result, if a theory is scale-invariant/superconformal, then there is a $U(1)_R$ symmetry, and $S_{\dot\alpha}$ is semi-chiral
\begin{equation}\label{eq:semiChiralST}
    D_- S_{\dot\alpha} = (\bm{Q} + \bar\partial)S_{\dot\alpha} = 0\,.
\end{equation}
Oftentimes, a theory is classically scale invariant and has a $U(1)_R$ symmetry, but this is broken by loop corrections\footnote{There is an apparent tension because corrections to the $\beta$-function persist beyond one-loop, but the triangle anomaly for the $R$-symmetry is one-loop exact. The resolution to this is neatly explained in \cite{Yonekura:2010mc}.}
\begin{equation}
    T_{\mu}^{\mu} 
        \propto \beta(g^2) \Tr F\wedge \star F\,
        \quad\text{and}\quad
    \partial_\mu j_R^\mu
        \propto C(R) \Tr F\wedge F\,.
\end{equation}
In terms of superfields, this is a Konishi anomaly \eqref{eq:veryKonishi}. Consequently, we expect one-loop corrections to $\bm{Q}$ to lift the semi-chiral superfield $S_{\dot\alpha}$ from cohomology when classical scaling symmetry is broken.

An even stronger result called ``holomorphic confinement'' occurs in pure SYM \cite{Budzik:2023xbr}. Pure SYM is classically scale invariant and has a $U(1)_R$ symmetry, but both of these are broken by one-loop corrections. Consequently, $S_{\dot\alpha}$ is in the classical $Q_0$-cohomology, but only $\partial^{\dot\alpha} S_{\dot\alpha}$ is expected in the quantum cohomology. As it turns out, $\partial^{\dot\alpha} S_{\dot\alpha}$ is not just $Q_1$-closed, but $Q_1$-exact, leaving the holomorphic twist of pure SYM completely topological.

\subsubsection{Anomalies as Higher Operations}
Obtaining corrections to the action of $\bm{Q}$ from anomalies is made even more precise in the holomorphic twist. The twisted theory describing the semi-chiral subsector can be considered as a local QFT in its own right, with form-valued superfields and a stress-tensor. A profitable viewpoint is to regard the theory as one where anti-holomorphic translations are gauge symmetries, having set\footnote{Or adding $\bm{Q}$ to $\bm{Q}_{\mathrm{BRST}}$ if gauge symmetries were already present.}
\begin{equation}
    \bm{Q}_{\mathrm{BRST}} := \bm{Q}\,,
\end{equation}
then Wess-Zumino consistency conditions demand $\bm{Q}_{\mathrm{BRST}}^2 = 0$. Computing anomalous corrections to $\bm{Q}$ is understood as solving the Wess-Zumino consistency conditions order-by-order in perturbation theory, preventing a ``holomorphic anomaly.''

To this end, we consider a cohomologically holomorphic theory $\calT_0$ (usually, but not necessarily, free) -- obtained as the twist of our original SQFT -- with nilpotent BRST charge $\bm{Q}_{\free}$. Turning on interactions in the original SQFT deforms the twisted theory $\calT_0$ by an interaction $\calI = g^i \calI_i$, i.e.\footnote{This process should be compared to how we compute anomalous scaling dimensions to an operator in conformal perturbation theory. Indeed, they are directly related, as recently emphasized by \cite{Scheinpflug:2025sqn}.}
\begin{equation}
    S_{\calT_0} \rightsquigarrow S_{\calT_0} + g^i \int_{\bbC^2} \dd^2z \, \calI^i\,.
\end{equation}
%In this case, the action of $\bm{Q}$ on operators in the deformed theory is given by 
We then study how the interaction deforms the semi-chiral ring of the free-theory.\footnote{This is an example of homotopy transfer, see Appendix D of \cite{Budzik:2023xbr}.} The action of the quantum corrected differential in the deformed theory on the $Q_{\text{free}}$-cohomology is given by
\begin{equation}\label{eq:QonO}
    \bm{Q}\! \calO = \{\calI, \calO\}_0 + \frac{\hbar}{2!}\{\calI,\calI,\calO\}_0 + \dots\,.
\end{equation}

The brackets $\{\,\cdot\,,\,\dots\,,\cdot\}_0$ are the ``higher operations'' of the original free cohomological field theory (defined below), and encode anomalies for the twisted theory in the form of (generalized) OPEs. For example, the $2$-ary bracket of two superfields is
\begin{equation}\label{eq:2aryProduct}
    \{\calO_1,\calO_2\}_0 = \oint_{S^3} \frac{\dd^2 z}{(2\pi i)^2} \calO_1(z)\!\calO_2(0)\,,
\end{equation}
where we use Wick contractions to produce a pole which is extracted by the $S^3$ contour integral. More generally, we can extract higher order poles or derivatives by integrating against elements $\rho \in H^{0,\bullet}(\bbC^2-0)$ in \eqref{eq:2aryProduct}; this can produce higher dimensional analogues of Virasoro and Kac-Moody symmetries \cite{faonte2019higher, Gwilliam:2018lpo, Saberi:2019fkq, Bomans:2023mkd, Budzik:2023xbr, Williams:2024mcc, Scheinpflug:2024mtn, Chen:2025ujx, Bomans:2025klo}.

These brackets generalize the OPE in two important ways. First, the inputs are necessarily form-valued fields from the cohomological theory. As previously emphasized, the OPE of $\bm{Q}$-closed operators is regular, so superfields are required for the integral \eqref{eq:2aryProduct} to be both valid and non-trivial. Thus this $2$-ary product includes the ``descent'' or ``secondary'' products of the theory, see e.g. \cite{Witten:1992yj,Lian:1992mn, Getzler:1994yd, Beem:2018fng, Oh:2019mcg, Garner:2022its}. Second, there are interesting higher $k$-ary operations $\{\,\cdot\,,\,\dots\,,\cdot\}_0$ which encode the failure of associativity of the OPE.\footnote{Readers familiar with higher-group symmetries in QFT will recognize the Jacobiator as a triple bracket which encodes the failure of the Jacobi identity in Lie 2-algebras.} We could already anticipate the need for these higher products when we noted that the OPE of two $\bm{Q}$-closed operators is $\bm{Q}$-exact, and thus invisible to $\bm{Q}$-cohomology. Together, the brackets form the structure of a (shifted) $L_\infty$-algebra, and \eqref{eq:QonO} is the appropriate Maurer-Cartan equation ensuring ``flatness'' of $\bm{Q}$ \cite{Gaiotto:2015aoa, Gaiotto:2024gii}. On $\bbC^2$, the precise derivation of \eqref{eq:QonO} in perturbation theory shows that the $(\ell+2)$-ary operation corresponds to the $\ell$-loop correction to the supercharge, i.e.
\begin{equation}
    \{\calI,\,\cdot\,\}_0 = Q_{0}\,,\quad
    \frac{1}{2!}\{\calI,\calI,\,\cdot\,\}_0 = Q_{1}\,,\quad
    \text{etc.}
\end{equation}
The Maurer-Cartan equation $\bm{Q}^2=0$ applied order by order imposes 
\begin{align}
    \sum_{n_1+n_2=n} Q_{n_1} Q_{n_2} = 0 \, .
\end{align}
For example,
\begin{align}
    \qty{Q_0,Q_1} = 0 \, , \qquad Q_1^2 + \qty{ Q_0 , Q_2} = 0 \, .
\end{align}
In particular, $Q_1$ does not square to zero on the $Q_\text{free}$-cohomology, but does square to zero on the $Q_0$-cohomology. In practice, we compute the corrected cohomology step by step using homotopy transfer (see Appendix D of \cite{Budzik:2023xbr}): first we compute $Q_0$-cohomology on $Q_{\text{free}}$-cohomology, then $Q_1$-cohomology on $Q_0$-cohomology, and so on.

For the purposes of this paper, we will define the (even more general) higher $\lambda$-brackets of an $L_\infty$-conformal algebra by
\begin{equation}\label{eq:BracketDefn}
    \{\calO_1 {}_{\lambda_1} \cdots \calO_k {}_{\lambda_k} \calO_{k+1}\}_0
        := \sum_{\Gamma} \int_{\bbC^{2k}} \!\!\left[\,\prod_{v=1}^k \frac{\dd^2 z_v}{(2\pi i)^{2}} e^{\lambda_v \cdot z_v}\right] Q_{\text{free}} W_\Gamma[\calO_1(z_1)\cdots \calO_k(z_v)\!\calO_{k+1}(0)]\,.
\end{equation}
Here, $\sum_{\Gamma}$ is a sum over Feynman diagrams, $W_{\Gamma}$ denotes the appropriate Wick contractions in $\calT_0$, and  $\lambda_v$ are holomorphic momenta flowing in at the vertex $v$, satisfying $\sum_{v=1}^k \lambda_v =: -\lambda_{k+1}$, which act as a generating function for the positive modes i.e. more singular terms of the (generalized) OPEs. In practice, formula \eqref{eq:BracketDefn} is derived by ``pulling down operators'' from the interaction term when treating a theory in perturbation theory, and the anomalous corrections to $\bm{Q}$ come from composite operator renormalization introducing extra terms which do not vanish as regulators go to $0$, i.e. they are literally perturbative anomalies \cite{Gaiotto:2024gii}. As we will discuss in the next section, the collection of Feynman diagrams contributing to \eqref{eq:BracketDefn} are highly restricted and computable.

The $L_\infty$-brackets, and thus action of $Q_n$ on local operators, are generically scheme dependent beyond first-order in perturbation theory. In principle, one may find a scheme where the brackets (and thus $Q_n$) vanish at some order in perturbation theory; this can depend sensitively on the particular theory (e.g., gauge groups and matter content). However, our computational procedure described here leaves integrands manifestly UV and IR finite and reproduces many known non-renormalization theorems in holomorphic-topological theories \cite{Kontsevich:1997vb, Budzik:2022mpd, Gaiotto:2024gii, Balduf:2024wwp}. The equivalence between different schemes, with different higher-brackets, can be expressed by finding an explicit quasi-isomorphism of the associated $L_\infty$-conformal algebras \cite{Kontsevich:1997vb, Budzik:2023xbr, Gaiotto:2024gii}.

\subsection{Computing Loop Corrections in the Holomorphic Twist}\label{sec:ComputingLoopCorrections}
Our interest concerns cohomological holomorphic field theories on $\bbC^2$, obtained by holomorphically twisting an SQFT, written in a first-order formulation
\begin{equation}\label{eq:FirstOrderAction}
    S = \int_{\bbC^2} \dd^2 z\, \left[(\Phi, \bar{\partial} \Phi) + \calI(\Phi)\right]\,.
\end{equation}
$\calI$ is a general interaction term, obtained by twisting an interaction of the original SQFT (see Section \ref{sec:Lagrangians}). $\Phi$ is a master $(0,\bullet)$-superfield containing all the fields $\phi^a$ of the holomorphic theory
\begin{equation}
    \Phi := \sum_{a} \phi^a v_a\,,\quad v_a \in V\,,
\end{equation}
and $V$ is an auxiliary super vector space whose elements $v_a$ are used to combine all fields $\phi^a$ together into $\Phi$.\footnote{This trick to organize all the fields into one master field will be familiar to string field theory aficionados.}

The superspace propagator is obtained by inverting the equation of motion
\begin{equation}
    \bar\partial \phi = 0\,.
\end{equation}
The inverse is the Bochner-Martinelli kernel, the $(0,1)$-form\footnote{Going forward, we write $\dot-,\dot+ = 1,2$ for ease of notation.}
\begin{equation}
    P(z,\bar{z}) := \frac{\bar{z}^2 \dd \bar{z}^{1} - \bar{z}^1 \dd\bar{z}^2}{|z|^4}\,,
\end{equation}
satisfying
\begin{equation}
    \bar\partial P(z,\bar{z}) = \pi^2 \delta^{(4)}(z) \,\dd^2 \bar{z}\,.
\end{equation}
The non-degenerate pairing $(\,\cdot\,,\,\cdot\,)$ on $V$ gives the matrix element $\eta_{ab} = (v_a,v_b)$. Thus the final propagator is
\begin{equation}\label{eq:propagatorFields}
    P^{ab}(z_{12},\bar{z}_{12}) := \expval*{\phi^a(z_1,\bar{z}_1) \phi^b(z_2,\bar{z}_2)} = \eta^{ab}\, P(z_{12},\bar{z}_{12})\,.
\end{equation}
Since the propagator is a $(0,1)$-form, the overall sign of the products depends sensitively on the order of multiplication, and it is important to consider graphs as directed graphs with a fixed edge labelling \cite{Budzik:2022mpd, Wang:2024sqm}.

In the previous section, we argued that perturbative corrections to $\bm{Q}$ in the SQFT are given by a sum over generalized Konishi anomaly/OPE brackets in the twisted theory, as in \eqref{eq:QonO}. As shown in \eqref{eq:BracketDefn}, these (perturbative) brackets are obtained by contracting (fields in) operators in the free theory. Thus, in principle, all that's left to do is (very carefully) perform Wick contractions on our form-valued fields, find the resulting graphs appearing in \eqref{eq:BracketDefn}, and compute the resulting Feynman integrals. The goal of the remainder of this section will be to give an operational description of this procedure before applying it to general Lagrangian SQFTs.

\subsubsection*{Wick Contracting Superfields} 
The entries to a bracket in \eqref{eq:BracketDefn} are superfields $\calO$, described in perturbation theory as (linear combinations of) normal-ordered products of free semichiral superfields $\phi^a$ and their holomorphic derivatives $\partial_{1}^{n_{1}} \partial_{2}^{n_{2}} \phi^a$. In other words, $\calO$ is a word of the form
\begin{equation}\label{eq:generalWord}
    \calO(0,0) \, =\,\, :\!\prod_{i=1}^{m} \partial_{1}^{n^{i}_{1}} \partial_{2}^{n^{i}_{2}} \phi^{a_i}(0,0)\!:\,.
\end{equation}
Normal-ordering is defined by subtracting all self-contractions of fields, it is associative and supercommutative on letters. Crucially, symmetry operators of the \textit{free} theory act by the Leibniz rule on normal ordered products of fields; this is obviously not true in an interacting theory, as emphasized by the familiar example of scaling dimensions $D$.\footnote{As explained in footnote 28 of \cite{Budzik:2023xbr}, one cannot generically find a composite operator renormalization scheme which is associative in the constituent operators and the interacting $\bm{Q}$ acts by Leibniz rule. The ability to find such a scheme is effectively a one-loop non-renormalization theorem.}

It is helpful to collect the derivatives of fields $\phi^a$ appearing in \eqref{eq:generalWord} in a generating function in a shifted position $w$, i.e. we write
\begin{equation}
    \phi^a(w,0) = \sum_{m,n=0}^\infty \frac{(w^1)^{m}(w^2)^{n}}{m! n!} \partial_{1}^{m} \partial_{2}^{n} \phi^{a}(0,0)\,. \label{eq:gen}
\end{equation}
Then, we can write propagators with a holomorphic shift along each edge in the graph
\begin{equation}\label{eq:shiftedPropagator}
    \expval*{\phi^a(z_1+w_1,\bar{z}_1) \phi^b(z_2+w_2,\bar{z}_2)} = P^{ab}(z_{12}+w_{12},\bar{z}_{12})\,,
\end{equation}
while integrating over the vertex positions $z_i$ in expressions like \eqref{eq:BracketDefn}. Considering these $w$-shifted propagators in Feynman diagrams lets us easily extract the action of $\bm{Q}$ on operators with derivatives from a single master integral formula. We note that the shift $w_{v,i}$ could be different for fields $\phi^a$ at the same vertex $\calO_v$, representing arbitrarily complicated derivatives in words \eqref{eq:generalWord}.

As always in perturbation theory, we should now perform all Wick contractions over fields inside the operators $\calO_v$ in \eqref{eq:BracketDefn}, replacing pairs of fields with the propagator \eqref{eq:shiftedPropagator} in the string
\begin{equation}
    \calO_1(z_1,\bar{z}_1)\cdots \calO_k(z_k,\bar{z}_k)\!\calO_{k+1}(0,0)\,.
\end{equation}
Note: many legs will generically be uncontracted; this is sensible since the brackets compute perturbative (i.e. local) anomalies, and thus are maps from the space of interactions/local operators to the space of local operators. Now we can consider the effect of $Q_{\free}$ on our partially contracted string
\begin{equation}
    W_{\Gamma}[\calO_1(z_1,\bar{z}_1)\cdots \calO_k(z_k,\bar{z}_k)\!\calO_{k+1}(0,0)]\,.
\end{equation}
The operator $Q_{\mathrm{free}}$ acts on the uncontracted superfields by the Leibniz rule on letters. Since all letters in \eqref{eq:generalWord} are semi-chiral superfields, $Q_{\free}$ acts by $-\bar\partial$, which can be integrated by parts to act on the propagators $P^{ab}(z+w,\bar{z})$ within the graph $\Gamma$, i.e.
\begin{equation}
   Q_{\free} W_{\Gamma}[\calO_1(z_1,\bar{z}_1)\cdots \calO_k(z_k,\bar{z}_k)\calO_{k+1}(0,0)]
        \mapsto \bar\partial\left[\,\prod_{e\in \Gamma_1} P^{ab}(z_e+w_e,\bar{z}_e)\right]\,.
\end{equation}
Applying $\bar\partial$ gives a sum over all ways to ``cut'' one of the propagators in $\Gamma$. Naively, the cut propagator $\bar\partial P^{ab}$ will give $0$ (modulo contact singularities from the $\delta$-function), and the action of $\bm{Q}$ reduces to its classical answer with Leibniz rule. Quantum mechanically, the integral over internal graph vertices needs to be regulated, as we review next, which leads to the anomalies and thus the modified action of $\bm{Q}$. We also note that the integrals are generally weighted by some external holomorphic momenta $e^{\lambda_n\cdot z_n}$ which acts as a generating functional for less singular terms in the OPE.

Before discussing the graphs, the integrals, and their regularization, let us note an important computational fact: the integrals required to compute the action of $\bm{Q}$ on general operators are universal, not depending on theory dependent combinatorial data such as $\eta^{ab}$. Given some uncontracted shifted (to include derivatives) superfields $\phi^a$ within an operator $\calO$, we can move the fields to the origin, leaving it completely independent of the integration variable $z$
\begin{equation}
    \calO(z,0) \,\,\, \supset \quad :\,\,\prod_{\mathclap{\mathrm{uncontracted}}}\,\, e^{z \cdot \partial_{w_i}}\,\phi^{a_i}(w_i,0)\,:\,.
\end{equation}
These factors $e^{z\cdot \partial_{w_i}}$ can then be combined with the $e^{\lambda_n\cdot z_n}$ terms, turning a general integral $I_{\Gamma}[\lambda;w]$ into a theory-independent differential operator, describing the contribution of the graph $\Gamma$ to $\bm{Q}$, and acting by
\begin{equation}\label{eq:IAsDifferentialOperator}
    \{\calO_1 {}_{\lambda_1} \cdots \calO_k {}_{\lambda_k} \calO_{k+1}\}_0\bigg\vert_{\Gamma}
        = I_{\Gamma}[\lambda_v + {\textstyle\sum_{i=1}^{m'}} \partial_{w_{v,i}}; w_e] \prod_{e \in \Gamma_1} \eta^{a_e b_e} \prod_{v \in \Gamma_0} \prod_{{\mathrm{un.}}} \phi^{a_{v,i}}(w_{v,i},0)\,.
\end{equation}

\subsubsection*{Laman Graphs, Operatope, and Bootstrap}
Quantum mechanically, the expression \eqref{eq:BracketDefn} requires regularization. In order to regulate the graph integrals, we rewrite the Bochner-Martinelli kernel in a Schwinger parametrization $t$, with UV cutoffs on the Schwinger times $t \in [\epsilon,\infty)$, replacing all propagators in $\Gamma$ by
\begin{equation}
    P(z+w,\bar{z}) \rightsquigarrow P_\epsilon(z+w,\bar{z},t)\,.
\end{equation}
This leads to non-trivial integrals and the anomalous action of $\bm{Q}$.

It is possible to define a combined propagator $\calP(z,\bar{z};t)$ which evaluates to $P_\epsilon$ or $\bar\partial P_\epsilon$ on $1$ and $0$ cycles respectively in the real projective space of Schwinger times $\mathbb{RP}^{|\Gamma_1|-1}$ (see \cite{Budzik:2022mpd, Gaiotto:2024gii} for details).\footnote{With a judicious change of variables, mimicking Fulton-MacPherson-Kontsevich compactification techniques for the Schwinger variables $t_e$, the integral can be brought to a manifestly UV (and IR finite form). Consequently, the holomorphic theories are finite and do not require counterterms. Even more surprisingly, manipulating graphs reveals that all loop integrals in theories with $\geq \! 2$ topological directions vanish \cite{Gaiotto:2024gii, Balduf:2024wwp}, providing a non-renormalization theorem for general holomorphic-topological QFTs.} With such a rewriting, we can recast the master integral $I_{\Gamma}[\lambda;w]$ in the form:
\begin{equation}\label{eq:MasterIntegral}
    I_\Gamma[\lambda;w] 
        = \int_{\bbC^{2k} \times \bbR \mathbb{P}_>^{|\Gamma_1|-1}} 
        \left[\,\prod^{k}_{v=1} \frac{\dd^2 z_v}{(2\pi i)^2} e^{\lambda_v \cdot z_v}\right] \left[\prod_{e \in \Gamma_1} \calP_{\epsilon}(z_{e(0)}-z_{e(1)}+w_e;t_e) \right]\,.
\end{equation}

\begin{figure}
\centering
\setlength\tabcolsep{12pt} % Default value: 6
\renewcommand{\arraystretch}{3} % Default value: 1
\begin{tabular}{ccc}
   \begin{tikzpicture}
        [
    	baseline={(current bounding box.center)},
    	line join=round
    	]
    	% Coordinates of the vertices of the graph
    	\coordinate (pd1) at (1.*\gS,0.*\gS);
    	\coordinate (pd2) at (-1.*\gS,0.*\gS);
    	% Label the vertices
    	\draw (pd1) node[GraphNode] {} ;
    	\draw (pd2) node[GraphNode] {};
    	% Draw the edges
        \draw[GraphEdge] (pd1) -- (pd2);
    \end{tikzpicture}&
   %%%%%%%%%%%%%%%%% second laman %%%%%%%%%
    \begin{tikzpicture}
		[
		baseline={(current bounding box.center)},
		line join=round
		]
		% Coordinates of the vertices of the graph
		\coordinate (pd1) at (-0.866*\gS,-0.5*\gS);
		\coordinate (pd2) at (0.866*\gS,-0.5*\gS);
		\coordinate (pd3) at (0.*\gS,1.*\gS);
		% Label the vertices
		\draw (pd1) node[GraphNode] {};
		\draw (pd2) node[GraphNode] {};
		\draw (pd3) node[GraphNode] {};
		% Draw the edges
		\draw[GraphEdge] (pd1) -- (pd2);
		\draw[GraphEdge] (pd1) -- (pd3);
		\draw[GraphEdge] (pd2) -- (pd3);
	\end{tikzpicture}&
	%%%%%%%%%%%%%% third laman %%%%%%%%%%%%%%%
	\begin{tikzpicture}
    	[
    	baseline={(current bounding box.center)},
    	line join=round
    	]
        \def\gS{1.5}
    	% Coordinates of the vertices of the graph
    	\coordinate (pd1) at (1.8671*\gS,0.435*\gS);
    	\coordinate (pd2) at (0.9338*\gS,0.87*\gS);
    	\coordinate (pd3) at (0.9339*\gS,0.*\gS);
    	\coordinate (pd4) at (0.*\gS,0.435*\gS);
    	% Label the vertices
    	\draw (pd1) node[GraphNode] {} ;
    	\draw (pd2) node[GraphNode] {}  ;
    	\draw (pd3) node[GraphNode] {}  ;
    	\draw (pd4) node[GraphNode] {}  ;
    	% Draw the edges
    	\draw[GraphEdge] (pd1) -- (pd2)  ;
    	\draw[GraphEdge] (pd1) -- (pd3)  ;
    	\draw[GraphEdge] (pd2) -- (pd3)  ;
    	\draw[GraphEdge] (pd2) -- (pd4)  ;
    	\draw[GraphEdge] (pd3) -- (pd4)  ;
	\end{tikzpicture}\\
	%%%%%%%%%%%%%%% fourth laman %%%%%%%%%%%%%%%
	\begin{tikzpicture}
		[
    	baseline={(current bounding box.center)},
	    line join=round
	    ]
        \def\gS{1.5};
    	% Coordinates of the vertices of the graph
    	\coordinate (pd1) at (1.7738*\gS,0.4892*\gS);
    	\coordinate (pd2) at (0.8297*\gS,0.*\gS);
    	\coordinate (pd3) at (0.0002*\gS,0.8472*\gS);
    	\coordinate (pd4) at (0.*\gS,0.1329*\gS);
    	\coordinate (pd5) at (0.8314*\gS,0.9802*\gS);
    	% Label the vertices
    	\draw (pd1) node[GraphNode] {} ;
    	\draw (pd2) node[GraphNode] {} ;
    	\draw (pd3) node[GraphNode] {} ;
    	\draw (pd4) node[GraphNode] {} ;
    	\draw (pd5) node[GraphNode] {} ;
    	% Draw the edges
    	\draw[GraphEdge] (pd1) -- (pd2) ;
    	\draw[GraphEdge] (pd1) -- (pd5) ;
    	\draw[GraphEdge] (pd2) -- (pd3);
    	\draw[GraphEdge] (pd2) -- (pd4);
    	\draw[GraphEdge] (pd3) -- (pd4) ;
    	\draw[GraphEdge] (pd3) -- (pd5) ;
    	\draw[GraphEdge] (pd4) -- (pd5) ;
	\end{tikzpicture}&
    \begin{tikzpicture}
        [
    	baseline={(current bounding box.center)},
    	line join=round
    	]
        \def\gS{1.5};
    	% Coordinates of the vertices of the graph
    	\coordinate (pd1) at (2.1987*\gS,0.0003*\gS);
    	\coordinate (pd2) at (1.627*\gS,0.7711*\gS);
    	\coordinate (pd3) at (1.0995*\gS,0.0413*\gS);
    	\coordinate (pd4) at (0.5706*\gS,0.7707*\gS);
    	\coordinate (pd5) at (0.*\gS,0.*\gS);
    
    	% Label the vertices
    	\draw (pd1) node[GraphNode] {};
    	\draw (pd2) node[GraphNode] {} ;
    	\draw (pd3) node[GraphNode] {} ;
    	\draw (pd4) node[GraphNode] {} ;
    	\draw (pd5) node[GraphNode] {} ;
    
    	% Draw the edges
    	\draw[GraphEdge] (pd1) -- (pd2) ;
    	\draw[GraphEdge] (pd1) -- (pd3);
    	\draw[GraphEdge] (pd2) -- (pd3) ;
    	\draw[GraphEdge] (pd2) -- (pd4) ;
    	\draw[GraphEdge] (pd3) -- (pd4) ;
    	\draw[GraphEdge] (pd3) -- (pd5) ;
    	\draw[GraphEdge] (pd4) -- (pd5) ;
    \end{tikzpicture}&
    %%%%%%%%%%%%%% sixth laman %%%%%%%%%%%%%%%%%%%%%%
    \begin{tikzpicture}
        [
    	baseline={(current bounding box.center)},
    	line join=round
    	]
        \def\gS{1.5};
    	% Coordinates of the vertices of the graph
    	\coordinate (pd1) at (0.0009*\gS,1.2196*\gS);
    	\coordinate (pd2) at (0.7208*\gS,0.9517*\gS);
    	\coordinate (pd3) at (0.*\gS,0.*\gS);
    	\coordinate (pd4) at (0.7203*\gS,0.2677*\gS);
    	\coordinate (pd5) at (1.6524*\gS,0.6091*\gS);
    	% Label the vertices
    	\draw (pd1) node[GraphNode] {};
    	\draw (pd2) node[GraphNode] {} ;
    	\draw (pd3) node[GraphNode] {} ;
    	\draw (pd4) node[GraphNode] {} ;
    	\draw (pd5) node[GraphNode] {} ;
    	% Draw the edges
    	\draw[GraphEdge] (pd1) -- (pd2) ;
    	\draw[GraphEdge] (pd1) -- (pd4) ;
    	\draw[GraphEdge] (pd2) -- (pd3) ;
    	\draw[GraphEdge] (pd2) -- (pd5) ;
    	\draw[GraphEdge] (pd3) -- (pd4) ;
    	\draw[GraphEdge] (pd4) -- (pd5) ;
    	\draw[GraphEdge] (pd2) -- (pd4) ;
    \end{tikzpicture}
\end{tabular}
    \caption{Laman graphs are the only graphs which can appear in the computation of brackets $\{\,\cdot\,,\dots,\,\cdot\,\}_0$ in a theory with action \eqref{eq:FirstOrderAction}. Even then, the contributions of some of these graphs to the action of $\bm{Q}$ may be ruled out by the interaction $\calI$. The propagator, the one-loop, the two-loop, and 3 three-loop Laman graphs are drawn above.}
    \label{fig:sixLamanGraphs}
\end{figure}
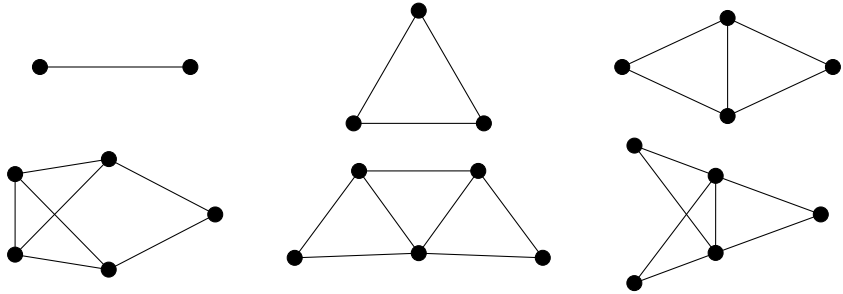

It was shown in \cite{Budzik:2022mpd} (see also \cite{Wang:2024sqm, Gaiotto:2024gii, Wang:2024tjf, Wang:2025rmu}) that the relevant graphs contributing (in principle) to the computation of \textit{any} $\ell$-loop, i.e. $(\ell+2)$-entry, $\lambda$-bracket are necessarily the $\ell$-loop Laman graphs \cite{henneberg1911graphische, pollaczek1927gliederung, laman1970graphs} (or $2$-Laman graphs in the language of arbitrary HT twists). This essentially follows from a careful study of the form degree of integrands in \eqref{eq:MasterIntegral}. The first few Laman graphs are depicted in Figure \ref{fig:sixLamanGraphs}. Crucially, at tree-level we have the propagator, and at one-loop there is only the triangle diagram.

The propagator master integral can be easily obtained by direct evaluation:
\begin{equation}
    \calI_{\prop}[\lambda;z] = e^{-\lambda_1 \cdot z_{12}}\,.
\end{equation}
The one loop-master integral \eqref{eq:MasterIntegral} was computed in \cite{Budzik:2022mpd}, and we repeat the computation for completeness in Appendix \ref{app:integral}. The result is:
\begin{align}
    \mathcal{I}_{\tri} [\lambda;z] &= -e^{-\lambda_1\cdot z_{13}-\lambda_2\cdot z_{23}} (\lambda_1\wedge \lambda_2) \qty[ \frac{1}{(\lambda_1\cdot Z)(\lambda_2\cdot Z)} + \frac{e^{-\lambda_1\cdot Z}}{(\lambda_1\cdot Z)(\lambda_3\cdot Z)} + \frac{e^{\lambda_2\cdot Z}}{(\lambda_2\cdot Z)(\lambda_3\cdot Z)} ] \, .
\end{align}
where $Z=z_{12}+z_{23}-z_{13}$ and $\lambda_1+\lambda_2+\lambda_3=0$ (for more details we refer to the appendix). Since we wish to convert this to a differential operator, as in \eqref{eq:IAsDifferentialOperator}, it is helpful to expand in the $z$ variables
\begin{align}
    \mathcal{I}_{\tri}[\lambda;z] &= e^{-\lambda_1\cdot z_{13}-\lambda_2\cdot z_{23}} (\lambda_1\wedge\lambda_2)\sum_{n=0}^\infty \frac{1}{(n+1)!}\frac{ (\lambda_2\cdot Z)^n - (-\lambda_1\cdot Z)^n }{(\lambda_1+\lambda_2)\cdot Z} \\
    &= e^{-\lambda_1\cdot z_{13}-\lambda_2\cdot z_{23}} (\lambda_1\wedge\lambda_2)  \sum_{n=0}^\infty \sum_{m=0}^\infty \frac{1}{(n+m+2)!} (-\lambda_1\cdot Z)^n (\lambda_2\cdot Z)^m \, . \label{eq:I}
\end{align}
Let us also note
\begin{align}
    \mathcal{I}_{\tri}[\lambda;0] &= \frac{1}{2}\lambda_1\wedge \lambda_2 \, .
\end{align}

In practice, we do not need to compute \eqref{eq:MasterIntegral} beyond the one-loop order. The Wess-Zumino consistency condition $\bm{Q}_{\mathrm{BRST}}^2 = 0$ implies a series of nested relations on the brackets when expanded in terms of \eqref{eq:QonO}, generalizing the Jacobi identity for our $L_\infty$ conformal algebra. Then, realizing the contribution of a graph to the action of $\bm{Q}$ as a differential operator, as in \eqref{eq:IAsDifferentialOperator}, we obtain a series of integral identities on graphs and their associated integrals, which can be solved order-by-order to bootstrap higher-loop master integrals, at least up to three-loops \cite{Budzik:2022mpd}.

\section{Lagrangian SUSY Gauge Theories}
\label{sec:Lagrangians}
In this section, we consider four-dimensional $\mathcal{N}=1$ Lagrangian theories with gauge Lie algebra $\mathfrak{g}$, chiral multiplets in a representation $R$ of $\mathfrak{g}$, and a chiral superpotential $W$. The holomorphic twist of the $\mathcal{N}=1$ vector multiplet $(A_{\mu},\lambda_\alpha)$ gives rise to a holomorphic $bc$ system, and the chiral multiplet $(\phi, \psi_\alpha)$ gives rise to a holomorphic $\beta\gamma$ system \cite{Johansen:1994aw, Costello:2011np,Elliott:2020ecf}. The resulting twisted theory is therefore a 4d holomorphic BF theory coupled to $\beta\gamma$ systems with a holomorphic superpotential $W[\gamma]$. 

The BV action takes the form
\begin{align}
    \int_{\mathbb{C}^2} \dd^2 z\, \Tr b \qty(\bar\partial c + \frac{1}{2} [c,c]) + \beta_I \qty(\bar\partial \gamma^I + c\gamma^I) + W[\gamma]  \, , \label{eq:L}
\end{align}
where the BV fields are collected into superfields 
\begin{alignat}{4}
   & b 
    &&= b^{(0)}+b^{(1)}+b^{(2)} \in \Omega^{0,\bullet}(\mathbb{C}^2,\mathfrak{g}^\vee) \, , \quad 
    && c 
    &&= c^{(0)}+c^{(1)}+c^{(2)} \in \Omega^{0,\bullet}(\mathbb{C}^2,\mathfrak{g})[1]\,, \\
   & 
   \beta_I 
   &&= \beta_I^{(0)}+\beta_I^{(1)}+\beta_I^{(2)} \in \Omega^{0,\bullet}(\mathbb{C}^2,R^\vee)[1] \, , \quad 
   && \gamma^I 
   &&= \gamma^{I(0)}+\gamma^{I(1)}+\gamma^{I(2)} \in \Omega^{0,\bullet}(\mathbb{C}^2,R) \, .
\end{alignat}
The tree-level differential is
\begin{alignat}{4}
    & Q_0 c &&= \frac{1}{2}[c,c] \, , \qquad  && Q_0 b &&= [c,b] - \mu(\beta,\gamma) \\
    & Q_0 \gamma^I &&= c\gamma^I \, , \qquad  &&  Q_0 \beta_I &&= \beta_I c + \partial_{\gamma^I} W \, ,
\end{alignat}
where $\mu(\beta,\gamma)$ is the moment map generating the symmetry gauged by the $bc$ system.

The tree-level BRST cohomology of the twisted theory can be mapped to the classical supercharge cohomology by \cite{Saberi:2019ghy} 
\begin{align}
    \gamma \sim \bar{\phi} \, , \qquad \beta \sim \psi_{+} \, , \qquad \partial_{\dot\alpha} c \sim \bar\lambda_{\dot\alpha} \, , \qquad b\sim F_{++} \, , \qquad \partial_{\dot\alpha} \sim D_{+\dot\alpha} \, .
\end{align}

In the next section, we will present the one-loop differential which governs the quantum corrections to the above tree-level cohomology. Then, we apply these results to pure $\mathcal{N}=1$ SYM (Section \ref{sec:N1}), $\mathcal{N}=2$ SYM (Section \ref{sec:N2}) and $\mathcal{N}=4$ SYM (Section \ref{sec:N4}). In these cases, all fields are in the adjoint representation and the one-loop differential admits compact local expressions in terms of superfields, see \eqref{eq:Q1N1}, \eqref{eq:Q1N2}, and \eqref{eq:Q1N4}. 

\subsection{One-Loop Supercharge}
\label{sec:Q1computation}
Now we turn to compute the one-loop differential in theories with first-order BV action of the form \eqref{eq:L}. 

As can be seen from the triangle diagrams in Section \ref{sec:ComputingLoopCorrections} and Appendix \ref{app:integral}, the one-loop differential $Q_1$ acts on a pair of fields and outputs a pair of uncontracted fields coming from the cubic interaction vertices. For operators of length $n>2$, the action of $Q_1$ is obtained by summing over its action on every distinct pair of fields $f_if_j$:
\begin{align}
    Q_1 ( f_1 \cdots f_n ) = \sum_{1\leq i<j\leq n} (-1)^{\epsilon_{ij}}\, Q_1(f_i f_j) f_1 \cdots f_{i-1} f_{i+1} \cdots f_{j-1} f_{j+1} \cdots f_n \, ,
\end{align}
with appropriate signs arising from the anticommutation of fermionic fields:
\begin{align}
    \epsilon_{ij} = |f_i|\sum_{k<i}|f_k| + |f_j|\sum_{\substack{k<j\\ k\ne i}}|f_k| \, ,
\end{align}
where $|f_i|\in\mathbb{Z}_2$ denotes the fermion parity of $f_i$.

We express the action of $Q_1$ on fields and their derivatives in a compact form using the generating function trick \eqref{eq:gen}. The action of $Q_1$ on fields with derivatives can then be obtained by expanding formulas as power series in the shift variables. The explicit formulas for the action of $Q_1$ on fields with derivatives are listed in Appendix \ref{app:derivatives}.

All one-loop actions are proportional to the triangle integral $\mathcal{I}_{\tri}[\lambda;z]$. As explained in Section \ref{sec:ComputingLoopCorrections}, the integral $\mathcal{I}_{\tri}[\lambda;z]$ is a power series in $\lambda_{\dot\alpha}$, which keeps track of derivatives on the uncontracted fields output from the bracket. We use the following notation to denote the differential operator form of the master integral $\mathcal{I}_{\tri}[\lambda;0,-w,-z]$ (i.e. with shifts $z_{12}=0,z_{23}=-w,z_{13}=-z$) acting on the two uncontracted fields coming from the interaction vertices:
\begin{align}
    \mathcal{D}_{w,z}^{\tri}(f,g) :=  \sum_{m,n=0}^\infty \frac{1}{(n+m+2)!} e^{z\cdot\partial} ((w-z)\cdot\partial)^n \partial_{\dot\alpha} f(y) \,  e^{w\cdot\partial} ((z-w)\cdot\partial)^m \partial^{\dot\alpha} g(y) \, \bigg|_{y=0},
\end{align}
where $z\cdot\partial = z^1\partial_{y_1}+z^2\partial_{y_2}$, i.e. the derivatives differentiate only the fields $f,g$ and not the shifts $z,w$. We also note:
\begin{align}
    \mathcal{D}^{\tri}_{0,0}(f,g) = \frac{1}{2} \partial_{\dot\alpha} f \partial^{\dot\alpha}g \, .
\end{align}

The action of $Q_1$ on pairs of fields that do not appear below is zero, i.e.:
\begin{align}
      Q_1(c c) = Q_1( \gamma \gamma) = Q_1(\gamma c) = Q_1(\beta c) = 0 \, .
\end{align}

\subsubsection*{Vector Multiplet}
\label{subsec:vector}
The twist of the vector multiplet is described by the holomorphic $bc$ system. The one-loop corrections to the supercharge in the 4d holomorphic BF theory were computed in \cite{Budzik:2023xbr}. Here we review them for completeness.

We use conventions for the adjoint representation of the Lie algebra $\mathfrak{g}$ so that generators are normalized with respect to the Killing form, i.e. the components can be raised and lowered using $\delta^{AB}$:
\begin{align}
    [t_A,t_B] = {f_{AB}}^{C} t_C \, , \qquad b^A=\delta^{AB}b_B \, , \qquad f_{ABC}= \delta_{CD} {f_{AB}}^D \, .
\end{align}
%where $\kappa$ is minus the dual Coxeter number.

There are two triangle diagrams involving $b$ and $c$:
\begin{center}
\begin{tikzpicture}
    [
    baseline={(current bounding box.center)},
    line join=round
    ]
    % Coordinates of the vertices of the graph
    \coordinate (pd1) at (-0.866*\gS,-0.5*\gS);
    \coordinate (pd2) at (0.866*\gS,-0.5*\gS);
    \coordinate (pd3) at (0.*\gS,1.*\gS);
    % Label the vertices
    \draw (pd1) node[GraphNode] {};
    \draw (pd2) node[GraphNode] {};
    \draw (pd3) node[GraphNode] {};
    % Draw the edges
    \draw[GraphEdge] (pd1) -- (pd2);
    \draw[GraphEdge] (pd1) -- (pd3);
    \draw[GraphEdge] (pd2) -- (pd3);
    % Draw letters
    \draw[] (-0.866*\gS-.2,-0.5*\gS-.5) node{$\frac{1}{2} \Tr b[c,c]$};
    \draw[] (0.866*\gS+.2,-0.5*\gS-.5) node{$\frac{1}{2} \Tr b[c,c]$};
    \draw[] (0.*\gS+.2,1.*\gS+.3) node{$b^A(z)c^B(w)$};
    %\node[] at (7,0) {$= \kappa^2 f_{ACD}f_{BCE} \qty[(\mathcal{I}[\lambda;0,-w,-z](c^Dc^E) + (z,D)\leftrightarrow (w,E)]$ };
\end{tikzpicture} \qquad\qquad
\begin{tikzpicture}
    [
    baseline={(current bounding box.center)},
    line join=round
    ]
    % Coordinates of the vertices of the graph
    \coordinate (pd1) at (-0.866*\gS,-0.5*\gS);
    \coordinate (pd2) at (0.866*\gS,-0.5*\gS);
    \coordinate (pd3) at (0.*\gS,1.*\gS);
    % Label the vertices
    \draw (pd1) node[GraphNode] {};
    \draw (pd2) node[GraphNode] {};
    \draw (pd3) node[GraphNode] {};
    % Draw the edges
    \draw[GraphEdge] (pd1) -- (pd2);
    \draw[GraphEdge] (pd1) -- (pd3);
    \draw[GraphEdge] (pd2) -- (pd3);
    % Draw letters
    \draw[] (-0.866*\gS-.2,-0.5*\gS-.5) node{$\frac{1}{2} \Tr b[c,c]$};
    \draw[] (0.866*\gS+.2,-0.5*\gS-.5) node{$\frac{1}{2} \Tr b[c,c]$};
    \draw[] (0.*\gS+.2,1.*\gS+.3) node{$b^A(z)b^B(w)$};
    %\node[] at (7,0) {$= \kappa^2 f_{ACD}f_{BCE} \qty[(\mathcal{I}[\lambda;0,-w,-z](-b^D c^E + c^D b^E) + (z,D)\leftrightarrow (w,E)]$ };
\end{tikzpicture}
\end{center}
Respectively, they contribute:
\begin{align}
     Q_1 (b^A (z) c^B (w) ) 
        &=  f_{ACD} f_{BCE} \, \mathcal{D}_{w,z}^{\tri}(c^D,c^E)\,, \\
     %=& \kappa^2 f_{ACD}f_{BCE} \qty( \mathcal{I}[\lambda;0,-w,-z](c^D,c^E) + (z,D)\leftrightarrow(w,E)  ) \\
     Q_1 (b^A (z) b^B (w) ) 
        &=  f_{ACD}f_{BCE} \left[ \mathcal{D}_{w,z}^{\tri}(c^D, b^E) -\mathcal{D}_{w,z}^{\tri} (b^D, c^E) \right] \, . \label{eq:Q1bzbw}
     %=& \kappa^2 f_{ACD}f_{BCE} \qty( -\mathcal{I}[\lambda;0,-w,-z] (b^D, c^E) + \mathcal{I}[\lambda;0,-w,-z](c^D, b^E) ) + (z,D)\leftrightarrow(w,E)   ) \\
\end{align}

\subsubsection*{Vector + Chiral Multiplet}
The twist of the chiral multiplet in a representation $R$ is described by a holomorphic $\beta\gamma$ system. We use the following convention: 
\begin{equation}
    \beta_I c\gamma^I = (\beta_I)_i c^A (T_A)^i_j (\gamma^I)^j \, .
\end{equation}
Here, $i,j,\dots$ are $R$ representation indices and $I,J,\dots$ are the flavour indices.

There are two triangle diagrams involving $\beta$ and $\gamma$:
\begin{center}
\begin{tikzpicture}
    [
    baseline={(current bounding box.center)},
    line join=round
    ]
    % Coordinates of the vertices of the graph
    \coordinate (pd1) at (-0.866*\gS,-0.5*\gS);
    \coordinate (pd2) at (0.866*\gS,-0.5*\gS);
    \coordinate (pd3) at (0.*\gS,1.*\gS);
    % Label the vertices
    \draw (pd1) node[GraphNode] {};
    \draw (pd2) node[GraphNode] {};
    \draw (pd3) node[GraphNode] {};
    % Draw the edges
    \draw[GraphEdge] (pd1) -- (pd2);
    \draw[GraphEdge] (pd1) -- (pd3);
    \draw[GraphEdge] (pd2) -- (pd3);
    % Draw letters
    \draw[] (-0.866*\gS-.2,-0.5*\gS-.5) node{$\beta_L c \gamma^L $};
    \draw[] (0.866*\gS+.2,-0.5*\gS-.5) node{$\beta_K c \gamma^K$};
    \draw[] (0.*\gS+.2,1.*\gS+.3) node{$(\beta_I)_i(z) (\gamma^J)^j(w)$};
\end{tikzpicture}\qquad\qquad
\begin{tikzpicture}
    [
    baseline={(current bounding box.center)},
    line join=round
    ]
    % Coordinates of the vertices of the graph
    \coordinate (pd1) at (-0.866*\gS,-0.5*\gS);
    \coordinate (pd2) at (0.866*\gS,-0.5*\gS);
    \coordinate (pd3) at (0.*\gS,1.*\gS);
    % Label the vertices
    \draw (pd1) node[GraphNode] {};
    \draw (pd2) node[GraphNode] {};
    \draw (pd3) node[GraphNode] {};
    % Draw the edges
    \draw[GraphEdge] (pd1) -- (pd2);
    \draw[GraphEdge] (pd1) -- (pd3);
    \draw[GraphEdge] (pd2) -- (pd3);
    % Draw letters
    \draw[] (-0.866*\gS-.2,-0.5*\gS-.5) node{$ W[\gamma] $};
    \draw[] (0.866*\gS+.2,-0.5*\gS-.5) node{$\beta_K c \gamma^K$};
    \draw[] (0.*\gS+.2,1.*\gS+.3) node{$(\beta_I)_i(z) (\beta_J)_j(w)$};
\end{tikzpicture} \qquad\qquad
\end{center}
\noindent They contribute:
\begin{align}
         Q_1 ((\beta_I)_i (z) (\gamma^J)^j (w) ) 
            =& - \delta_I^J (T_B T_A)^j_i \mathcal{D}_{w,z}^{\tri}(c^A, c^B) \,,\\
         %&=  - 2\delta_I^J\mathcal{I}[\lambda;0,-z,-w](c^A, c^B) (T_A T_B)^j_i \\
         % \delta_I^J \qty( -\mathcal{I}[\lambda;0,-w,-z](c^B, c^A) - \mathcal{I}[\lambda;0,-z,-w](c^Ac^B)) (T_A T_B)^j_i \\
         \begin{split}
         Q_1 ((\beta_I)_i (z) (\beta_J)_j (w) ) 
            =& - (T_A)_i^k \mathcal{D}_{w,z}^{\tri}( c^A , \partial_{\gamma^{Ik}} \partial_{\gamma^{Jj}} W ) -  (T_A)^k_j \mathcal{D}_{w,z}^{\tri}( \partial_{\gamma^{Ii}}\partial_{\gamma^{Jk}}W , c^A )  \,. 
            %=& -\mathcal{D}_{w,z}^{\tri}( c^A , \partial_{\gamma^{Ik}} \partial_{\gamma^{Jj}} W ) (T_A)_i^k + \mathcal{D}_{z,w}^{\tri}( c^A ,\partial_{\gamma^{Ii}}\partial_{\gamma^{Jk}}W ) (T_A)^k_j \\
            %& - \mathcal{D}_{w,z}^{\tri}( \partial_{\gamma^{Ii}}\partial_{\gamma^{Jk}}W , c^A ) (T_A)^k_j + \mathcal{D}_{z,w}^{\tri}( \partial_{\gamma^{Ik}}\partial_{\gamma^{Jj}} W , c^A ) (T_A)^k_i\,.
         \end{split}
\end{align}

The new interaction also gives a contribution to the $Q_1(bb)$ action:
\begin{center}
\begin{tikzpicture}
    [
    baseline={(current bounding box.center)},
    line join=round
    ]
    % Coordinates of the vertices of the graph
    \coordinate (pd1) at (-0.866*\gS,-0.5*\gS);
    \coordinate (pd2) at (0.866*\gS,-0.5*\gS);
    \coordinate (pd3) at (0.*\gS,1.*\gS);
    % Label the vertices
    \draw (pd1) node[GraphNode] {};
    \draw (pd2) node[GraphNode] {};
    \draw (pd3) node[GraphNode] {};
    % Draw the edges
    \draw[GraphEdge] (pd1) -- (pd2);
    \draw[GraphEdge] (pd1) -- (pd3);
    \draw[GraphEdge] (pd2) -- (pd3);
    % Draw letters
    \draw[] (-0.866*\gS-.2,-0.5*\gS-.5) node{$\beta_J c \gamma^J$};
    \draw[] (0.866*\gS+.2,-0.5*\gS-.5) node{$\beta_I c \gamma^I$};
    \draw[] (0.*\gS+.2,1.*\gS+.3) node{$b^A(z)b^B(w)$};
    %\node[] at (7,0) {$= \kappa^2 f_{ACD}f_{BCE} \qty[(\mathcal{I}[\lambda;0,-w,-z](c^Dc^E) + (z,D)\leftrightarrow (w,E)]$ };
\end{tikzpicture} 
\end{center}

This modifies \eqref{eq:Q1bzbw} to
\begin{align}
    \begin{split}
     Q_1 (b^A (z) b^B (w) ) 
        &= f_{ACD}f_{BCE} \left[ \mathcal{D}_{w,z}^{\tri}(c^D, b^E) -\mathcal{D}_{w,z}^{\tri} (b^D, c^E)  \right] \\
        &\hphantom{=}+ (T_BT_A)^i_j \mathcal{D}_{w,z}^{\tri}((\gamma^{I})^j,(\beta_I)_i)  -  (T_AT_B)^i_j \mathcal{D}_{w,z}^{\tri}((\beta_{I})_i,(\gamma^I)^j)  \, .
    \end{split}
\end{align}

We also get five additional diagrams: 
\begin{center}
\begin{tikzpicture}
    [
    baseline={(current bounding box.center)},
    line join=round
    ]
    % Coordinates of the vertices of the graph
    \coordinate (pd1) at (-0.866*\gS,-0.5*\gS);
    \coordinate (pd2) at (0.866*\gS,-0.5*\gS);
    \coordinate (pd3) at (0.*\gS,1.*\gS);
    % Label the vertices
    \draw (pd1) node[GraphNode] {};
    \draw (pd2) node[GraphNode] {};
    \draw (pd3) node[GraphNode] {};
    % Draw the edges
    \draw[GraphEdge] (pd1) -- (pd2);
    \draw[GraphEdge] (pd1) -- (pd3);
    \draw[GraphEdge] (pd2) -- (pd3);
    % Draw letters
    \draw[] (-0.866*\gS-.2,-0.5*\gS-.5) node{$\beta_J c \gamma^J $};
    \draw[] (0.866*\gS+.2,-0.5*\gS-.5) node{$\frac{1}{2}\Tr b[c,c]$};
    \draw[] (0.*\gS+.2,1.*\gS+.3) node{$b^A(z) (\gamma^I)^i(w)$};
\end{tikzpicture}\qquad\qquad
\begin{tikzpicture}
    [
    baseline={(current bounding box.center)},
    line join=round
    ]
    % Coordinates of the vertices of the graph
    \coordinate (pd1) at (-0.866*\gS,-0.5*\gS);
    \coordinate (pd2) at (0.866*\gS,-0.5*\gS);
    \coordinate (pd3) at (0.*\gS,1.*\gS);
    % Label the vertices
    \draw (pd1) node[GraphNode] {};
    \draw (pd2) node[GraphNode] {};
    \draw (pd3) node[GraphNode] {};
    % Draw the edges
    \draw[GraphEdge] (pd1) -- (pd2);
    \draw[GraphEdge] (pd1) -- (pd3);
    \draw[GraphEdge] (pd2) -- (pd3);
    % Draw letters
    \draw[] (-0.866*\gS-.2,-0.5*\gS-.5) node{$\beta_K c \gamma^K $};
    \draw[] (0.866*\gS+.2,-0.5*\gS-.5) node{$\beta_J c \gamma^J $};
    \draw[] (0.*\gS+.2,1.*\gS+.3) node{$b^A(z) (\gamma^I)^i(w)$};
\end{tikzpicture}
\end{center}
\begin{center}
\begin{tikzpicture}
    [
    baseline={(current bounding box.center)},
    line join=round
    ]
    % Coordinates of the vertices of the graph
    \coordinate (pd1) at (-0.866*\gS,-0.5*\gS);
    \coordinate (pd2) at (0.866*\gS,-0.5*\gS);
    \coordinate (pd3) at (0.*\gS,1.*\gS);
    % Label the vertices
    \draw (pd1) node[GraphNode] {};
    \draw (pd2) node[GraphNode] {};
    \draw (pd3) node[GraphNode] {};
    % Draw the edges
    \draw[GraphEdge] (pd1) -- (pd2);
    \draw[GraphEdge] (pd1) -- (pd3);
    \draw[GraphEdge] (pd2) -- (pd3);
    % Draw letters
    \draw[] (-0.866*\gS-.2,-0.5*\gS-.5) node{$ \beta_J c \gamma^J $};
    \draw[] (0.866*\gS+.2,-0.5*\gS-.5) node{$\frac{1}{2}\Tr b[c,c]$};
    \draw[] (0.*\gS+.2,1.*\gS+.3) node{$(\beta_I)_i(z) b^A(w) $};
\end{tikzpicture} \qquad\qquad
\begin{tikzpicture}
    [
    baseline={(current bounding box.center)},
    line join=round
    ]
    % Coordinates of the vertices of the graph
    \coordinate (pd1) at (-0.866*\gS,-0.5*\gS);
    \coordinate (pd2) at (0.866*\gS,-0.5*\gS);
    \coordinate (pd3) at (0.*\gS,1.*\gS);
    % Label the vertices
    \draw (pd1) node[GraphNode] {};
    \draw (pd2) node[GraphNode] {};
    \draw (pd3) node[GraphNode] {};
    % Draw the edges
    \draw[GraphEdge] (pd1) -- (pd2);
    \draw[GraphEdge] (pd1) -- (pd3);
    \draw[GraphEdge] (pd2) -- (pd3);
    % Draw letters
    \draw[] (-0.866*\gS-.2,-0.5*\gS-.5) node{$ \beta_K c \gamma^K $};
    \draw[] (0.866*\gS+.2,-0.5*\gS-.5) node{$\beta_J c\gamma^J$};
    \draw[] (0.*\gS+.2,1.*\gS+.3) node{$(\beta_I)_i(z) b^A(w) $};
\end{tikzpicture} \qquad\qquad
\begin{tikzpicture}
    [
    baseline={(current bounding box.center)},
    line join=round
    ]
    % Coordinates of the vertices of the graph
    \coordinate (pd1) at (-0.866*\gS,-0.5*\gS);
    \coordinate (pd2) at (0.866*\gS,-0.5*\gS);
    \coordinate (pd3) at (0.*\gS,1.*\gS);
    % Label the vertices
    \draw (pd1) node[GraphNode] {};
    \draw (pd2) node[GraphNode] {};
    \draw (pd3) node[GraphNode] {};
    % Draw the edges
    \draw[GraphEdge] (pd1) -- (pd2);
    \draw[GraphEdge] (pd1) -- (pd3);
    \draw[GraphEdge] (pd2) -- (pd3);
    % Draw letters
    \draw[] (-0.866*\gS-.2,-0.5*\gS-.5) node{$ W[\gamma] $};
    \draw[] (0.866*\gS+.2,-0.5*\gS-.5) node{$\beta_J c\gamma^J$};
    \draw[] (0.*\gS+.2,1.*\gS+.3) node{$(\beta_I)_i(z) b^A(w) $};
\end{tikzpicture}
\end{center}

They contribute
\begin{align}
    Q_1( b^A(z) (\gamma^I)^i(w) ) &= %\kappa f_{ABC} (T_B)^i_j [ I[\lambda;0,-w,-z](c^C, (\gamma^I)^j) - I[\lambda;0,-z,-w]((\gamma^I)^j, c^C) ]  \\
    %& +(T_BT_A)^i_j( -\mathcal{I}[\lambda;0,-w,-z]((\gamma^I)^j,c^B) +\mathcal{I}[\lambda;0,-z,-w](c^B,(\gamma^I)^j) )  \\
     f_{ABC} (T_B)^i_j \mathcal{D}_{w,z}^{\tri}(c^C, (\gamma^I)^j)   - (T_BT_A)^i_j \mathcal{D}_{w,z}^{\tri}((\gamma^I)^j,c^B)  \,, \\
     \begin{split}
    Q_1( (\beta_I)_i(z) b^A(w) ) &= %\kappa f_{ABC}(T_B)_i^j (- \mathcal{I}[\lambda;0,-w,-z]((\beta_I)_j,c^C)+\mathcal{I}[\lambda;0,-z,-w](c^C,(\beta_I)_j) ) \\
    %&+ (T_AT_B)^j_i ( \mathcal{I}[\lambda;0,-w,-z](c^B,(\beta_I)_j) + \mathcal{I}[\lambda;0,-z,-w]((\beta_I)_j), c^B)   \\
    %& + \mathcal{I}[\lambda;0,-w,-z]( \partial_{\gamma^{Ii}}\partial_{\gamma^{Jj}}W , (\gamma^J)^k ) (T_A)^j_k - \mathcal{I}[\lambda;0,-z,-w] ( (\gamma^J)^k , \partial_{\gamma^{Ii}}\partial_{\gamma^{Jj}} W ) (T_A)^j_k \\
     - f_{ABC}(T_B)_i^j \mathcal{D}_{w,z}^{\tri}((\beta_I)_j,c^C) +   (T_AT_B)^j_i  \mathcal{D}_{w,z}^{\tri}(c^B,(\beta_I)_j)  \\
    &\hphantom{=} +  (T_A)^j_k \mathcal{D}_{w,z}^{\tri}( \partial_{\gamma^{Ii}}\partial_{\gamma^{Jj}}W , (\gamma^J)^k )  \, . %- \mathcal{D}_{z,w}^{\tri} ( (\gamma^J)^k , \partial_{\gamma^{Ii}}\partial_{\gamma^{Jj}} W ) (T_A)^j_k \, .
    \end{split}
\end{align}

\subsubsection*{Vector + Adjoint Matter}
We summarize the results in the special case when $R$ is the adjoint representation, i.e. $(T_A)^{B}_C={f_{AC}}^B$:
\begin{align}
     Q_1 (b^A (z) c^B (w) ) 
        &= f_{ACD} f_{BCE} \mathcal{D}_{w,z}^{\tri}(c^D,c^E)\,, \\
    \begin{split}
     Q_1 (b^A (z) b^B (w) ) 
        &=  f_{ACD}f_{BCE} \left[ \mathcal{D}_{w,z}^{\tri}(c^D, b^E) -\mathcal{D}_{w,z}^{\tri} (b^D, c^E) \right. \\
        &\hphantom{=} \left.+\mathcal{D}_{w,z}^{\tri}((\beta_I)^D,(\gamma^I)^E) -\mathcal{D}_{w,z}^{\tri}((\gamma^I)^D,(\beta_I)^E) \right]\,, \label{eq:oneW}
    \end{split} \\
    Q_1 ((\beta_I)^A (z) (\gamma^J)^B (w) ) 
        &= \delta_I^J  f_{ACD}f_{BCE} \mathcal{D}_{w,z}^{\tri}(c^D,c^E)\,, \\
        \begin{split}
    Q_1 ((\beta_I)^A (z) (\beta_J)^B (w) ) 
        &= - \left[ f_{ACD} \mathcal{D}_{w,z}^{\tri}(c^D,\partial_{\gamma^{IC}} \partial_{\gamma^{JB}} W) \right. \\
        &\hphantom{=}\left. + f_{BCE} \mathcal{D}_{w,z}^{\tri}(\partial_{\gamma^{IA}}\partial_{\gamma^{JC}} W,c^E) \right] \,, \label{eq:twoW}
        \end{split} \\
    Q_1( b^A(z) (\gamma^I)^B(w) ) 
        &= f_{ACD} f_{BCE} \left[ \mathcal{D}_{w,z}^{\tri}(c^D, (\gamma^I)^E) -\mathcal{D}_{w,z}^{\tri}((\gamma^I)^D, c^E)\right]\,, \\
    \begin{split}
    Q_1( (\beta_I)^A(z) b^B(w)) 
        &= f_{ACD}f_{BCE} \left[\mathcal{D}_{w,z}^{\tri}((\beta_I)^D,c^E)+\mathcal{D}_{w,z}^{\tri}(c^D,(\beta_I)^E)\right] \\ 
        &\hphantom{=}- f_{BCE} \mathcal{D}_{w,z}^{\tri}(\partial_{\gamma^{IA}}\partial_{\gamma^{KC}} W,(\gamma^K)^E)\,. \label{eq:threeW}
    \end{split}
\end{align}
Specializing to $W[\gamma]=\Tr\gamma^1[\gamma^2,\gamma^3]$, which is the $\mathcal{N}=4$ SYM case, \eqref{eq:oneW}, \eqref{eq:twoW} and \eqref{eq:threeW} give:
\begin{align}
    Q_1 (b^A (z) b^B (w) ) 
        &= f_{ACD}f_{BCE} \left[ \mathcal{D}_{w,z}^{\tri}(c^D, b^E) -\mathcal{D}_{w,z}^{\tri} (b^D, c^E) \right. \\
        &\hphantom{=} \left.+\mathcal{D}_{w,z}^{\tri}((\beta_I)^D,(\gamma^I)^E) -\mathcal{D}_{w,z}^{\tri}((\gamma^I)^D,(\beta_I)^E) \right]\,, \\
    Q_1 ((\beta_I)^A (z) (\beta_J)^B (w) ) 
        &= \varepsilon_{IJK} f_{ACD}f_{BCE}\left[\mathcal{D}_{w,z}^{\tri}(c^D,(\gamma^K)^E)-\mathcal{D}_{w,z}^{\tri}((\gamma^K)^D,c^E) \right]\,, \\
    Q_1( (\beta_I)^A(z) b^B(w)) 
        &= f_{ACD}f_{BCE} \left[\mathcal{D}_{w,z}^{\tri}((\beta_I)^D,c^E)+\mathcal{D}_{w,z}^{\tri}(c^D,(\beta_I)^E)\right. \\ 
        &\hphantom{=}+ \left.\varepsilon_{IJK} \mathcal{D}_{w,z}^{\tri}((\gamma^J)^D,(\gamma^K)^E)\right]\,.
\end{align}

\subsection{\texorpdfstring{$\mathcal{N}=1$}{N=1} Super Yang-Mills}
\label{sec:N1}

The holomorphic twist of pure $\mathcal{N}=1$ SYM is the four-dimensional holomorphic BF theory \cite{Johansen:1994aw, Costello:2011np, Costello:2013zra, Elliott:2020ecf}. This example has been studied extensively in \cite{Budzik:2023xbr}, where, in particular the tree-level and one-loop cohomologies were computed for $SU(N)$ at infinite-$N$, and studied numerically at $N=2,3$.

The BV fields of the holomorphic BF theory can be conveniently combined into a single superfield valued in $\mathbb{C}[z^1,z^2,\theta]\otimes\mathfrak{g}$:
\begin{align}
    C = c + \theta b \, . \label{eq:N1C}
\end{align}
Then the BV action takes the form of a Chern-Simons theory on $\bbC^{2|1}$
\begin{align}
    \frac{1}{2}\int_{\mathbb{C}^{2|1}} \dd^2 z\, \dd\theta \, \Tr C\qty(\bar\partial C+\frac{1}{3}[C,C]) \, ,
\end{align}
and the tree-level differential can be written as
\begin{align}
    Q_0 C = \frac{1}{2}[C,C] \, .
\end{align}
This is the standard Lie algebra cohomology differential.

The one-loop differential $Q_1$ was computed in \cite{Budzik:2023xbr} and reviewed in Section \ref{subsec:vector}. We write it below for vanishing shifts $z=w=0$:\footnote{We thank Jiyoo Park for pointing out a missing factor of $1/2$ in formulas for $z=w=0$ in Sections \ref{sec:N1}, \ref{sec:N2}, \ref{sec:N4}.}
\begin{align}
     Q_1 (b^A c^B ) 
        &=  \frac{1}{2} f_{ACD}f_{BCE} \, \partial_{\dot\alpha} c^D\partial^{\dot\alpha} c^E \,, \\
     Q_1(b^A b^B) 
        &= \frac{1}{2} f_{ACD}f_{BCE} \left[ \partial_{\dot\alpha} c^D \partial^{\dot\alpha} b^E -\partial_{\dot\alpha} b^D\partial^{\dot\alpha} c^E  \right]\,.
\end{align}
We observe that the one-loop action can be equivalently expressed using the superfield \eqref{eq:N1C} as
\begin{equation}
    Q_1( C^A(\theta) C^B(\theta') ) = - \frac{1}{2} f_{ACD}f_{BCE} (\theta-\theta') \partial_{\dot\alpha} C^D(\theta) \partial^{\dot\alpha} C^E(\theta') \, . \label{eq:Q1N1}
\end{equation}
This is the simplest example where we can rewrite the one-loop differential using the superfield notation.

\subsection{\texorpdfstring{$\mathcal{N}=2$}{N=2} Super Yang-Mills}
\label{sec:N2}

The holomorphic twist of pure $\mathcal{N}=2$ SYM is the four-dimensional holomorphic BF theory coupled to a single adjoint $\beta\gamma$ system \cite{Costello:2011np, Elliott:2020ecf}.

In this case, the BV fields can be combined into two superfields valued in $\mathbb{C}[z_1,z_2,\theta]\otimes \mathfrak{g}$ and $\mathbb{C}[z_1,z_2,\theta]\otimes \mathfrak{g}^\vee$:
\begin{align}
    C= c+ \theta\gamma \, , \qquad 
    B  = \beta+\theta b \, . \label{eq:N2BC}
\end{align}
Then the BV action takes the form
\begin{align}
    \int_{\mathbb{C}^{2|1}} \dd^2 z \dd\theta \Tr B \qty( \bar\partial C + \frac{1}{2} [C,C] ) \, , 
\end{align}
and the tree-level differential can be written as
\begin{align}
    Q_0 C = \frac{1}{2}[C,C] \, , \qquad Q_0 B=[C,B] \, .
\end{align}

The one-loop differential computed in Section \ref{sec:Q1computation} for vanishing shifts $z=w=0$ is
\begin{align}
     Q_1 (b^A c^B ) 
        &=  \frac{1}{2} f_{ACD}f_{BCE} \, \partial_{\dot\alpha} c^D\partial^{\dot\alpha} c^E\,,  \\
     Q_1(b^A b^B) 
        &= \frac{1}{2} f_{ACD}f_{BCE} \left[\partial_{\dot\alpha} c^D \partial^{\dot\alpha} b^E -\partial_{\dot\alpha} b^D\partial^{\dot\alpha} c^E + \partial_{\dot\alpha}\beta^D\partial^{\dot\alpha}{\gamma}^E  - \partial_{\dot\alpha} {\gamma}^D\partial^{\dot\alpha}\beta^E \right] \,,\\
     Q_1 (\beta^A {\gamma}^B ) 
        &=  \frac{1}{2} f_{ACD}f_{BCE} \, \partial_{\dot\alpha} c^D\partial^{\dot\alpha} c^E\,, \\
    Q_1(b^A{\gamma}^B) 
        &= \frac{1}{2} f_{ACD}f_{BCE} \left[\partial_{\dot\alpha} c^D \partial^{\dot\alpha} {\gamma}^E-\partial_{\dot\alpha} \gamma^D \partial^{\dot\alpha} c^E\right]\,, \\
    Q_1(\beta^A b^B) 
        &= \frac{1}{2} f_{ACD}f_{BCE} \left[ \partial_{\dot\alpha} \beta^D \partial^{\dot\alpha} c^E + \partial_{\dot\alpha} c^D \partial^{\dot\alpha} \beta^E \right] \, .
\end{align}
We again find that the one-loop action can be recast using the superfields \eqref{eq:N2BC} as
\begin{align}
    Q_1(B^A(\theta)C^B(\theta')) 
        &= -\frac{1}{2} f_{ACD}f_{BCE} (\theta-\theta') \partial_{\dot\alpha} C^D(\theta) \partial^{\dot\alpha} C^E(\theta') \label{eq:Q1N2} \\
    Q_1( B^A(\theta) B^B(\theta') ) 
        &= -\frac{1}{2} f_{ACD}f_{BCE} (\theta-\theta') \left[\partial_{\dot\alpha} B^D(\theta) \partial^{\dot\alpha} C^E(\theta')+\partial_{\dot\alpha} C^D(\theta) \partial^{\dot\alpha} B^E(\theta')\right] \, .
    %Q_1( C^A(\theta) C^B(\theta') ) &= 0 \, . 
\end{align}

\subsection{\texorpdfstring{$\mathcal{N}=4$}{N=4} Super Yang-Mills}
\label{sec:N4}
Finally, we consider $\mathcal{N}=4$ SYM, in which case the twisted theory is the 4d holomorphic BF theory with three adjoint $\beta\gamma$ systems and the superpotential
\begin{align}
     W[\gamma] = \Tr \gamma^1[\gamma^2,\gamma^3] = \frac{1}{3!} \varepsilon_{IJK} f_{ABC} (\gamma^I)^A(\gamma^J)^B(\gamma^K)^C \, .
\end{align}
The tree-level differential is
\begin{alignat}{4}
    & Q_0 c 
        &&= \frac{1}{2}[c,c] \, , \qquad  
        && Q_0 b &&= [c,b] - [\gamma^I,\beta_I]\,, \\
    & Q_0 \gamma_I 
        &&= [c, \gamma_I] \, , \qquad  
        &&  Q_0 \beta_I 
        &&= [c,\beta_I] + \frac{1}{2}\varepsilon_{IJK}[\gamma^J,\gamma^K] \, .
\end{alignat}

The tree-level cohomology is isomorphic to the classical supercharge cohomology in the untwisted theory. For the choice of twisting supercharge $\bm{Q}:= Q^4_-$, the fields are identified by
\begin{align}
    \gamma^I \sim \Phi^{4I} \, , \qquad \beta_I \sim \Psi_{I+} \, , \qquad \partial_{\dot\alpha} c \sim \bar\Psi^4_{\dot\alpha} \, , \qquad b\sim F_{++} \, , \qquad \partial_{\dot\alpha} \sim D_{+\dot\alpha} \, ,
\end{align}
where $\Phi^{IJ}$, $I,J=1,\dots,4$ are the six real scalars, and $\Psi_{I\dot\alpha}$, $\bar\Psi^I_{\dot\alpha}$ are the four chiral and four anti-chiral fermions.

We again collect the BV fields into a superfield valued in $\mathbb{C}[z_1,z_2,\theta_1,\theta_2,\theta_3]\otimes \mathfrak{g}$ \cite{Chang:2013fba}:
\begin{align}
    C = c + \theta_I \gamma^I +\frac{1}{2}\varepsilon^{IJK}\theta_I\theta_J\beta_K+\theta_1\theta_2\theta_3 b \, . \label{eq:C}
\end{align}
The BV action takes the form of a Chern-Simons theory on $\bbC^{2|3}$
\begin{align}
    \frac{1}{2}\int_{\mathbb{C}^{2|3}} \dd^2 z
\, \dd^3 \theta \, \Tr C\qty(\bar\partial C + \frac{1}{3}[C,C] ) \, ,
\end{align}
and the action of the tree-level differential can be written as
\begin{align}
    Q_0 C=\frac{1}{2}[C,C] \, .
\end{align}

In case of $\mathcal{N}=4$ SYM, we can use the residual $\mathfrak{psl}(3|3)$ algebra, which is the subalgebra of the $\mathfrak{psl}(4|4)$ superconformal algebra that commutes with $\bm{Q}=Q^4_{-}$, to fix the one-loop corrections knowing only one term e.g. $Q_1(bc)$. 

The (tree-level\footnote{The tree-level action of $\mathfrak{psl}(3|3)$ might also receive loop corrections, however the one-loop differential $Q_1$ (anti)commutes with the tree-level generators $R_0$ of $\mathfrak{psl}(3|3)$ on $Q_0$-cohomology, i.e. $\qty{Q_1,R_0}+\qty{Q_0,R_1}=0$.}) action of the $\mathfrak{psl}(3|3)$ can be conveniently represented by super vector fields acting on the superspace \cite{Saberi:2019fkq}:
\begin{alignat}{4}
    & Q^I_{+} 
        && \sim \partial_{\theta_I} \, , \qquad 
        && S_I^{+} 
        &&\sim \theta_I \qty( z^{\dot\alpha}\partial_{\dot\alpha} + \theta_J\partial_{\theta_J} )\,, \\
    & \tilde{Q}_{I\dot \alpha} 
        && \sim \theta_I \partial_{\dot\alpha} \, , \qquad 
        && \tilde{S}^{I\dot\alpha} 
        &&\sim z^{\dot\alpha} \partial_{\theta_I} \, .
\end{alignat}
In particular, we will use
\begin{alignat}{5}
    & Q^I_+ : \quad 
        && c \mapsto \gamma^I \, , \quad 
        && \gamma^J \mapsto  -\varepsilon^{IJK}\beta_K \,\quad 
        && \beta_J \mapsto \delta^I_J b \, , \quad 
        && b\mapsto 0\,, \\
    & S_I^+ : \quad 
        && c \mapsto 0 \, , \quad 
        && \gamma^J \mapsto -\delta^J_I z^{\dot\alpha}\partial_{\dot\alpha}c \, , \quad 
        && \beta_J \mapsto -\varepsilon_{IJK} (1+z^{\dot\alpha}\partial_{\dot\alpha})\gamma^K \, , \quad && b\mapsto -(2+z_{\dot\alpha}\partial_{\dot\alpha}) \beta_I \, .
\end{alignat}

Now we can fix the loop corrections starting from $Q_1(bc)$:
\begin{align}
    Q_1(b\gamma^I) 
        &= - Q_+^I Q_1 (bc)\,, \\
    Q_1(b\beta_I) 
        &= \frac{1}{2} \varepsilon_{IJK} Q^J_+ Q_1 (b\gamma^K)\,, \\
    Q_1(bb) 
        &= - Q^I_+ Q_1 (b\beta_I)\,,  \\
    Q_1(\beta_I\gamma^J) 
        &= -\frac{1}{2} S^+_I Q_1 (b\gamma^J)\,, \\
    Q_1(\beta_I\beta_J) 
        &= \frac{1}{2} S^+_I Q_1(b\beta_J) - \frac{1}{2} \varepsilon_{IJK}Q_1(b\gamma^K) \,.
\end{align}
We arrive at
\begin{align}
     Q_1 (b^A c^B ) 
        &=  \frac{1}{2} f_{ACD}f_{BCE} \, \partial_{\dot\alpha} c^D\partial^{\dot\alpha} c^E \,, \\
    \begin{split}
     Q_1(b^A b^B) 
        &= \frac{1}{2} f_{ACD}f_{BCE} \left[ \partial_{\dot\alpha} c^D \partial^{\dot\alpha} b^E -\partial_{\dot\alpha} b^D\partial^{\dot\alpha} c^E \right.\\
        &\hphantom{=2\kappa^2 f_{ACD}f_{BCE}[]}\left. +\partial_{\dot\alpha}(\beta_I)^D\partial^{\dot\alpha}(\gamma^I)^E - \partial_{\dot\alpha} (\gamma^I)^D\partial^{\dot\alpha}(\beta_I)^E  \right]\,,
    \end{split}\\
     Q_1 ((\beta_I)^A (\gamma^J)^B ) 
        &=  \frac{1}{2} \delta^J_I f_{ACD}f_{BCE} \, \partial_{\dot\alpha} c^D\partial^{\dot\alpha} c^E \,,  \label{eq:oneterm} \\
     Q_1 ((\beta_I)^A (\beta_J)^B ) 
        &= \frac{1}{2} \varepsilon_{IJK} f_{ACD}f_{BCE} \left[ \partial_{\dot\alpha}c^D \partial^{\dot\alpha}(\gamma^K)^E -\partial_{\dot\alpha} (\gamma^K)^D \partial^{\dot\alpha} c^E \right] \,,\\
    Q_1(b^A(\gamma^I)^B) 
        &= \frac{1}{2} f_{ACD}f_{BCE} \left[\partial_{\dot\alpha} c^D \partial^{\dot\alpha} (\gamma^I)^E-\partial_{\dot\alpha} (\gamma^I)^D \partial^{\dot\alpha} c^E\right] \,, \\
    \begin{split}
    Q_1((\beta_I)^A b^B) 
        &= \frac{1}{2} f_{ACD}f_{BCE} \left[ \partial_{\dot\alpha} (\beta_I)^D \partial^{\dot\alpha} c^E + \partial_{\dot\alpha} c^D \partial^{\dot\alpha} (\beta_I)^E\right.\\
        &\hphantom{= 2\kappa^2 f_{ACD}f_{BCE}}\left.+\varepsilon_{IJK}\partial_{\dot\alpha} (\gamma^J)^D \partial^{\dot\alpha}(\gamma^K)^E \right] \, ,
    \end{split}
\end{align}
which matches the formulas in Section \ref{sec:Q1computation}. The formulas including derivatives can be found in Appendix \ref{app:derivatives}. The correction \eqref{eq:oneterm} was computed also in \cite{Choi:2025bhi}. 

Finally, we observe that the above one-loop differential can be written in a compact
superfield form analogous to the $\mathcal{N}=1$ and $\mathcal{N}=2$ cases:
\begin{align}
        Q_1 (C^A(\theta) C^B(\theta')) = -\frac{1}{2} f_{ACD}f_{BCE}  (\theta_1-\theta_1')(\theta_2-\theta_2')(\theta_3-\theta_3') \partial_{\dot\alpha} C^D(\theta) \partial^{\dot\alpha} C^E(\theta') \, . \label{eq:Q1N4}
\end{align}

\subsubsection{Loop Corrections in the Planar Limit}

Here, we discuss the one-loop corrections presented in the previous section in the planar limit of $\mathcal{N}=4$ SYM with gauge group $SU(N)$.\footnote{We thank Davide Gaiotto for discussions on this point.}

Due to the Lie algebra factors $f_{ACD}f_{BCE}$, the one-loop supercharge $Q_1$ can join and split traces. To illustrate this, write one term in the $Q_1$ action schematically as
\begin{align}
    Q_1(X^A Y^B) \sim f_{ACD}f_{BCE} U^D V^E \, .
\end{align}
For simplicity, $X,Y,U$ are bosonic and $V$ is fermionic. 

Now we consider the action of $Q_1$ on neighbouring and non-neighbouring letters $X,Y$ inside an example single-trace:
\begin{align}
    Q_1 \Tr (XYAB) &\sim -N\Tr(UVAB) - \Tr(UV)\Tr(AB) \\
    &\hphantom{\sim} + \Tr(U)\Tr(VAB) + \Tr(V)\Tr(UAB)  + \dots \\
    Q_1 \Tr (XAYB) &\sim \Tr(UA)\Tr(VB) + \Tr(UB)\Tr(VA) \\
    &\hphantom{\sim} - \Tr(A)\Tr(UVB) - \Tr(B)\Tr(UAV) + \dots \, ,
\end{align}
where the ellipses denote terms where $Q_1$ acts on other pairs of letters. We see that $Q_1$ acting on neighboring letters in a single-trace can produce a single-trace output and a multi-trace output.
%\footnote{Whether the single-trace to single-trace part is dominant in large-$N$ depends on the normalization of multi-traces.}
Similarly, $Q_1$ can join traces. Acting on $X$ in the first trace and $Y$ in the second trace gives
\begin{align}
    Q_1\qty( \Tr(XA) \Tr(YB) ) &\sim \Tr(UAVB) + \Tr(AUBV) \nonumber \\
    &\hphantom{\sim} -\Tr(UABV)-\Tr(AUVB)+ \dots \, .
\end{align}

We would like to study the action of $Q_1$ on the infinite-$N$ single-trace tree-level cohomology, which is in one-to-one correspondence with the $1/16$-BPS spectrum of $\mathcal{N}=4$ SYM. The cohomology classes, computed in \cite{Chang:2013fba}, can be represented by traces of symmetrized products of $\gamma^I$,
\begin{align}
    \Tr \gamma^{(I_1} \gamma^{I_2} \dots \gamma^{I_n)} \, , \label{eq:gammas}
\end{align}
and their $\mathfrak{psl}(3|3)$ descendants.

We first show that these classes are $Q_1$-closed at any $N$. It is straightforward to see that $Q_1(\gamma^I\gamma^J)=0$ since there are no triangle diagrams that could contribute to this term. Therefore, $Q_1$ vanishes on the representatives \eqref{eq:gammas}. Since $Q_1$ also (anti)commutes with the (tree-level) $\mathfrak{psl}(3|3)$ symmetry generators (on $Q_0$-cohomology), as discussed in the preceding section, we conclude that all $\mathfrak{psl}(3|3)$ descendants of \eqref{eq:gammas} are $Q_1$-closed. %The $Q_1$-closedness of \eqref{eq:gammas} and theirs descendants holds at any $N$.

At infinite-$N$, we can restrict the action of $Q_1$ to the part that maps single-traces to single-traces (as split into single- and multi-traces is well-defined at infinite-$N$). Since $Q_1$ vanishes on all infinite-$N$ single-trace cohomologies, the image of $Q_1$ is zero and no such classes can be $Q_1$-exact. We therefore conclude that the part of $Q_1$ action that maps single-traces to single-traces does not lift the infinite-$N$ tree-level single-trace cohomology classes.\footnote{In the pure SYM example, $Q_1$ acts non-trivially on the infinite-$N$ tree-level single-trace cohomology \cite{Budzik:2023xbr}.} Analyzing the part of $Q_1$ which can split and join traces is more subtle and we leave it for future work.

\acknowledgments
We would like to thank Davide Gaiotto and Xi Yin for useful discussions and comments on the draft. We also thank Chi-Ming Chang, Anirudh Deb, Alexandre Homrich, Eunwoo Lee, Ying-Hsuan Lin, Jiyoo Park, Surya Raghavendran, Matteo Sacchi, Aiden Suter, Pedro Vieira and Jingxiang Wu for helpful discussions. The work of JK is supported by the NSERC PDF program. The work of KB was supported by the Simons Collaboration on Celestial Holography.

\appendix

\section{The One-Loop Master Integral}
\label{app:integral}

For completeness, we reproduce the computation of the master integral \eqref{eq:MasterIntegral} associated to the triangle diagram \eqref{eq:triangle} from \cite{Budzik:2022mpd}.
\begin{equation}
\begin{tikzpicture}
    [
	baseline={(current bounding box.center)},
	line join=round
	]
    \def\gS{1.5};
	% Coordinates of the vertices of the graph
	\coordinate (pd1) at (-0.866*\gS,-0.5*\gS);
	\coordinate (pd2) at (0.*\gS,1.*\gS);
	\coordinate (pd3) at (0.866*\gS,-0.5*\gS);
	% Label the vertices
	\draw (pd1) node[GraphNode] {} node[left] {$\lambda_{2}$};
	\draw (pd2) node[GraphNode] {} node[above] {$\lambda_{3}$};
	\draw (pd3) node[GraphNode] {} node[right] {$\lambda_{1}$};
	% Draw the edges
	\draw[GraphEdge] (pd1) -- (pd2) node[midway, left] {$z_{23}$};
	\draw[GraphEdge] (pd1) -- (pd3) node[midway, above] {$z_{12}$};
	\draw[GraphEdge] (pd2) -- (pd3) node[midway, right] {$z_{13}$};
   \end{tikzpicture}\,.
\label{eq:triangle}
\end{equation}

%\begin{align}
%    Q_1 \mathcal{O} &= \frac{1}{2} \qty{ \mathcal{I} , \mathcal{I}, \mathcal{O}}
%\end{align}
We define the oriented loop variable $Z=z_{12}+z_{23}-z_{13}$ and the holomorphic momenta $\lambda_i$, satisfying $\lambda_1+\lambda_2+\lambda_3=0$. We also define $\lambda\wedge\lambda'=\lambda_1\lambda'_2-\lambda_2\lambda'_1$, $\lambda\cdot z = \lambda_1 z^1 + \lambda_2 z^2$, $|x|^2=|x_1|^2+|x_2|^2$, and the combined propagator on spacetime and the space of Schwinger parameters (see \cite{Budzik:2022mpd, Gaiotto:2024gii}):
\begin{equation}
     \calP(x,\bar{x};t) := \frac{\dd t}{t}(\bar{x}^2 \dd\bar{x}^1 - \bar{x}^1 \dd\bar{x}_2)K_t(x) + \dd^2\bar{x}\,  K_t(x)\,,
\end{equation}
where $K_t(x)$ is the heat-kernel $K_t(x) := t^{-2} e^{-|x|^2/t}$.

We start from the integral over spacetime and Schwinger parameters:
\begin{align}
    &\int_{\mathbb{C}^{4}\times \mathbb{RP}_{>}^2} \dd^2 x_1 \dd^2 x_2 \, e^{\lambda_1\cdot x_1+\lambda_2\cdot x_2 + \lambda_3\cdot x_3} \mathcal{P}(x_1\!-\!x_2\!+\!z_{12};t_1) \mathcal{P}(x_2\!-\!x_3\!+\!z_{23};t_2) \mathcal{P}(x_1\!-\!x_3\!+\!z_{13};t_3) \\
    &=e^{-\lambda_1\cdot z_{13}-\lambda_2\cdot z_{23}}\int_{\mathbb{C}^4\times\mathbb{RP}_>^2} \dd^2 x_1 \dd^2 x_2 \, e^{\lambda_1\cdot x_1+\lambda_2\cdot x_2} \mathcal{P}(x_1\!-\!x_2\!+\!Z;t_1) \mathcal{P}(x_2;t_2) \mathcal{P}(x_1;t_3) \\ 
    &=-e^{-\lambda_1\cdot z_{13}-\lambda_2\cdot z_{23}}\int_{\mathbb{C}^4\times\mathbb{RP}_>^2} \dd^2 x_1 \dd^2 x_2 \dd^2\bar{x}_1\dd^2\bar{x}_2 \qty(\frac{\dd t_1 \dd t_2}{t_1t_2} + \frac{\dd t_2\dd t_3}{t_2t_3} + \frac{\dd t_3\dd t_1}{t_1 t_3} ) \frac{1}{t_1^2 t_2^2 t_3^2} \\
    &\qquad\qquad \times e^{\lambda_1\cdot x_1+\lambda_2\cdot x_2} (\bar{x}_1^2\bar{x}_2^1-\bar{x}_1^1\bar{x}_2^2) e^{-\frac{1}{t_3}x_1\cdot\bar{x}_1-\frac{1}{t_2}x_2\cdot\bar{x}_2-\frac{1}{t_1}(x_1-x_2+Z)\cdot(\bar{x}_1-\bar{x}_2)}  \, .
\end{align}
The spacetime integrals are Gaussian, after evaluating them we get
\begin{align}
    &e^{-\lambda_1\cdot z_{13}-\lambda_2\cdot z_{23}} (\lambda_1\wedge\lambda_2) \int_{\mathbb{RP}_>^2} \qty(\frac{\dd t_1\dd t_2}{t_1t_2} + \frac{\dd t_2 \dd t_3}{t_2t_3} + \frac{\dd t_3\dd t_1}{t_1 t_3} ) \frac{t_1t_2t_3}{(t_1+t_2+t_3)^3} e^{-\frac{t_3\lambda_1\cdot Z-t_2\lambda_2\cdot Z}{t_1+t_2+t_3}} \, .
\end{align}
Next, we do the integral over the Schwinger parameters on the region $t_1+t_2+t_3=1$:
\begin{align}
    \int_{\mathbb{RP}_>^2} & \qty(t_3\dd t_1\dd t_2 + t_1\dd t_2 \dd t_3 + t_2\dd t_3\dd t_1 ) \, e^{-(t_3\lambda_1\cdot Z-t_2\lambda_2\cdot Z)} \\
    &= -\frac{1}{(\lambda_1\cdot Z)(\lambda_2\cdot Z)} + \frac{e^{-\lambda_1\cdot Z}}{(\lambda_1\cdot Z)((\lambda_1+\lambda_2)\cdot Z)}+\frac{e^{\lambda_2\cdot Z}}{(\lambda_2\cdot Z)((\lambda_1+\lambda_2)\cdot Z)} \, .
\end{align}
Together, we get
\begin{align}
\mathcal{I}_{\tri}[\lambda;z] = & \, -e^{-\lambda_1\cdot z_{13}-\lambda_2\cdot z_{23}} (\lambda_1\wedge\lambda_2) \\
&\times\qty[ \frac{1}{(\lambda_1\cdot Z)(\lambda_2\cdot Z)} - \frac{e^{-\lambda_1\cdot Z}}{(\lambda_1\cdot Z)((\lambda_1+\lambda_2)\cdot Z)}-\frac{e^{\lambda_2\cdot Z}}{(\lambda_2\cdot Z)((\lambda_1+\lambda_2)\cdot Z)} ] \, ,
\end{align}
which can also be written as
\begin{align}
  \mathcal{I}_{\tri}[\lambda;z] &=  -e^{-\lambda_1\cdot z_{13}-\lambda_2\cdot z_{23}} (\lambda_1\wedge\lambda_2) \qty[\frac{1}{(\lambda_1\cdot Z)(\lambda_2\cdot Z)} + \frac{e^{-\lambda_1\cdot Z}}{(\lambda_1\cdot Z)(\lambda_3\cdot Z)}+\frac{e^{\lambda_2\cdot Z}}{(\lambda_2\cdot Z)(\lambda_3\cdot Z)}] \, .
\end{align}

\section{One-Loop Supercharges on Derivatives}
\label{app:derivatives}
In this appendix, we present the explicit action of the one-loop supercharge $Q_1$ on fields with derivatives, in a form convenient for symbolic implementation e.g. in Mathematica.

In Section \ref{sec:Q1computation}, we computed the one-loop supercharge action on the generating functions \eqref{eq:gen}, using shifted Feynman diagrams \eqref{eq:triangle}. The generating functions encode all holomorphic derivatives and the action of $Q_1$ on fields with derivatives can be obtained by expanding both sides of formulas in Section \ref{sec:Q1computation} as a power series in the shifts $z=(z^1,z^2)$ and $w=(w^1,w^2)$.

Here, we present the explicit formulas for the action of the one-loop supercharge on expressions of the form $f \partial_1^m\partial_2^n g$, where $f,g$ are two elementary fields. Then, the action on derivatives of the first field can be obtained recursively by
\begin{align}
    Q_1 ( \partial_{\dot\alpha} f g ) = \partial_{\dot\alpha} Q_1 ( fg ) - Q_1 ( f \partial_{\dot\alpha} g ) \, , \label{eq:rec}
\end{align}
since $Q_1$ commutes with derivatives.

We expand
\begin{align}
    Q_1 ( f(0)g(w) ) = \sum_{m,n=0}^\infty \frac{1}{m!n!} (w^1)^m (w^2)^n Q_1( f \partial_1^m \partial_2^n g  ) \, .
\end{align}

To match the coefficients of $(w^1)^m(w^2)^n$, we expand the integrals \eqref{eq:I} in power series as
\begin{align}
    \mathcal{I}_{\tri}[\lambda; 0,-w,0]=(\lambda_1\wedge\lambda_2)\sum_{m,n=0}^\infty C_{mn} (w^1)^m(w^2)^n \, ,
\end{align}
where after many combinatorial manipulations (see the next appendix), one can show that
\begin{align}
  m!n! C_{mn} = \frac{1}{m+n+2} \sum_{k=0}^m \sum_{l=0}^n \frac{1}{k+l+1} \binom{m}{k} \binom{n}{l} (\lambda_1)_1^{k} (\lambda_1)_2^{l} (\lambda_2)_1^{m-k} (\lambda_2)_2^{n-l} \, . \label{eq:Cmn}
\end{align}
We use this to write
\begin{align}
    \begin{split}
    Q_1 ( b^A \partial_1^m \partial_2^n c^B ) 
        &= f_{ACD}f_{BCE} \frac{1}{m+n+2} \sum_{k=0}^m \sum_{l=0}^n \frac{1}{k+l+1} \binom{m}{k} \binom{n}{l} \\
        & \cdot\bigg[\partial_1^{k} \partial_2^{l} \partial_{\dot\alpha} c^D \partial_1^{m-k} \partial_2^{n-l} \partial^{\dot\alpha} c^E \bigg]\,,
    \end{split}\\
    Q_1 ( b^A \partial_1^m \partial_2^n b^B ) 
        &= f_{ACD}f_{BCE}\frac{1}{m+n+2} \sum_{k=0}^m \sum_{l=0}^n \frac{1}{k+l+1} \binom{m}{k} \binom{n}{l}\nonumber \\
        &\cdot\bigg[ \partial_1^{k} \partial_2^{l} \partial_{\dot\alpha} c^D \partial_1^{m-k} \partial_2^{n-l} \partial^{\dot\alpha} b^E -\partial_1^{k} \partial_2^{l} \partial_{\dot\alpha} b^D \partial_1^{m-k} \partial_2^{n-l} \partial^{\dot\alpha} c^E \\
        & +\partial_1^{k} \partial_2^{l} \partial_{\dot\alpha} (\beta_I)^D \partial_1^{m-k} \partial_2^{n-l} \partial^{\dot\alpha} (\gamma^I)^E -\partial_1^{k} \partial_2^{l} \partial_{\dot\alpha} (\gamma^I)^D \partial_1^{m-k} \partial_2^{n-l} \partial^{\dot\alpha} (\beta_I)^E  \bigg]\nonumber\,,
        \\
    \begin{split}
    Q_1 ( (\beta_I)^A \partial_1^m \partial_2^n (\gamma^J)^B ) 
        &= \delta^J_I f_{ACD}f_{BCE} \frac{1}{m+n+2} \sum_{k=0}^m \sum_{l=0}^n \frac{1}{k+l+1} \binom{m}{k} \binom{n}{l} \\
        & \cdot\bigg[\partial_1^{k} \partial_2^{l} \partial_{\dot\alpha} c^D \partial_1^{m-k} \partial_2^{n-l} \partial^{\dot\alpha} c^E \bigg]\,,
    \end{split}\\
    \begin{split}
    Q_1 ( (\beta_I)^A \partial_1^m \partial_2^n (\beta_J)^B )  
        &= \varepsilon_{IJK} f_{ACD}f_{BCE} \frac{1}{m+n+2} \sum_{k=0}^m \sum_{l=0}^n \frac{1}{k+l+1} \binom{m}{k} \binom{n}{l} \\
        & \cdot\bigg[\partial_1^{k} \partial_2^{l} \partial_{\dot\alpha} c^D \partial_1^{m-k} \partial_2^{n-l} \partial^{\dot\alpha} (\gamma^K)^E - \partial_1^{k} \partial_2^{l} \partial_{\dot\alpha} (\gamma^K)^D \partial_1^{m-k} \partial_2^{n-l} \partial^{\dot\alpha} c^E \bigg]\,,
    \end{split}\\
    \begin{split}
        Q_1 ( b^A \partial_1^m \partial_2^n (\gamma^I)^B ) &= f_{ACD}f_{BCE} \frac{1}{m+n+2} \sum_{k=0}^m \sum_{l=0}^n \frac{1}{k+l+1} \binom{m}{k} \binom{n}{l} \\
        & \cdot\bigg[\partial_1^{k} \partial_2^{l} \partial_{\dot\alpha} c^D \partial_1^{m-k} \partial_2^{n-l} \partial^{\dot\alpha} (\gamma^I)^E - \partial_1^{k} \partial_2^{l} \partial_{\dot\alpha} (\gamma^I)^D \partial_1^{m-k} \partial_2^{n-l} \partial^{\dot\alpha} c^E \bigg]\,,
    \end{split}\\
    Q_1 ( (\beta_I)^A \partial_1^m \partial_2^n b^B ) 
        &= f_{ACD}f_{BCE} \frac{1}{m+n+2} \sum_{k=0}^m \sum_{l=0}^n \frac{1}{k+l+1} \binom{m}{k} \binom{n}{l} \nonumber \\
        & \cdot\bigg[\partial_1^{k} \partial_2^{l} \partial_{\dot\alpha} (\beta_I)^D \partial_1^{m-k} \partial_2^{n-l} \partial^{\dot\alpha} c^E + \partial_1^{k} \partial_2^{l} \partial_{\dot\alpha} c^D \partial_1^{m-k} \partial_2^{n-l} \partial^{\dot\alpha} (\beta_I)^E \\
        &+ \varepsilon_{IJK} \partial_1^{k} \partial_2^{l} \partial_{\dot\alpha} (\gamma^J)^D \partial_1^{m-k} \partial_2^{n-l} \partial^{\dot\alpha} (\gamma^K)^E  \bigg] \, ,\nonumber
\end{align}
for adjoint matter with cubic superpotential $W[\gamma]=\Tr\gamma^1[\gamma^2,\gamma^3]$, and analogously for other representations.

\section{Combinatorial Manipulations}\label{app:CombinatorialIdentity}

We summarize the combinatorial manipulations needed to arrive at \eqref{eq:Cmn}. We expand the integral \eqref{eq:I}:
\begin{align}
    \mathcal{I}_{\tri}[\lambda;0,-w,0] 
        &= (\lambda_1\wedge \lambda_2) e^{\lambda_2\cdot w}\sum_{k=0}^\infty \sum_{j=0}^\infty \frac{1}{(k+j+2)!} (\lambda_1\cdot w)^k (-\lambda_2\cdot w)^j \\
    &= (\lambda_1\wedge \lambda_2) \sum_{p=0}^\infty \frac{1}{p!} \sum_{k=0}^\infty \sum_{j=0}^\infty (-1)^j \frac{1}{(k+j+2)!} (\lambda_1\cdot w)^k (\lambda_2\cdot w)^{j+p} \\
    &= (\lambda_1\wedge \lambda_2) \sum_{k=0}^\infty (\lambda_1\cdot w)^k \sum_{r=0}^\infty (\lambda_2\cdot w)^r \sum_{j=0}^r (-1)^j \frac{1}{(r-j)!} \frac{1}{(k+j+2)!} \\
    &= (\lambda_1\wedge \lambda_2) \sum_{k=0}^\infty (\lambda_1\cdot w)^k \sum_{r=0}^\infty (\lambda_2\cdot w)^r  \frac{1}{(r+k+2)r!(k+1)!} \, . \label{eq:lastline}
\end{align}
In the last line, we used the following combinatorial identity:
\begin{align}
    %&\sum_{m=0}^r (-1)^m \binom{r}{m} \frac{(m+l)!}{(m+l+k+n+2)!} = \frac{l!r!}{(l+r+n+k+2)!} \binom{r+n+k+1}{r} \\
    & \sum_{j=0}^r (-1)^j \frac{1}{(r-j)!} \frac{1}{(k+j+2)!} = \frac{1}{(r+k+2)r!(k+1)!} \, . \label{eq:combi}
\end{align}

We can group the terms in \eqref{eq:lastline} with the same powers of $(w_1)^m(w_2)^n$ as
\begin{align}
    %&(\lambda_1\wedge\lambda_2) \frac{1}{m+n+2}\sum_{k=0}^m\sum_{l=0}^n \frac{1}{(m+n-k-l)!} \frac{1}{(k+l+1)!} \binom{k+l}{l} (\lambda_{1})_1^k(\lambda_1)_2^l \binom{m+n-k-l}{n-l} (\lambda_2)_1^{m-k} (\lambda_2)_2^{n-l} \\
    (\lambda_1\wedge\lambda_2) \frac{1}{m+n+2}\sum_{k=0}^m\sum_{l=0}^n \frac{1}{k+l+1} \frac{1}{k!l!} (\lambda_{1})_1^k(\lambda_1)_2^l \frac{1}{(m-k)!(n-l)!} (\lambda_2)_1^{m-k} (\lambda_2)_2^{n-l} \, .
\end{align}

\bibliographystyle{JHEP}

\bibliography{mono}

\providecommand{\href}[2]{#2}\begingroup\raggedright\begin{thebibliography}{10}

\bibitem{Seiberg:1994bz}
N.~Seiberg, {\it {Exact results on the space of vacua of four-dimensional SUSY gauge theories}},  {\em Phys. Rev. D} {\bf 49} (1994) 6857--6863, [\href{http://arxiv.org/abs/hep-th/9402044}{{\tt hep-th/9402044}}].

\bibitem{CDSW}
F.~Cachazo, M.~R. Douglas, N.~Seiberg, and E.~Witten, {\it {Chiral rings and anomalies in supersymmetric gauge theory}},  {\em JHEP} {\bf 12} (2002) 071, [\href{http://arxiv.org/abs/hep-th/0211170}{{\tt hep-th/0211170}}].

\bibitem{Sohnius:1981sn}
M.~F. Sohnius and P.~C. West, {\it {Conformal Invariance in N=4 Supersymmetric Yang-Mills Theory}},  {\em Phys. Lett. B} {\bf 100} (1981) 245.

\bibitem{deWit:1984rvr}
B.~de~Wit, P.~G. Lauwers, and A.~Van~Proeyen, {\it {Lagrangians of N=2 Supergravity - Matter Systems}},  {\em Nucl. Phys. B} {\bf 255} (1985) 569--608.

\bibitem{Seiberg:1988ur}
N.~Seiberg, {\it {Supersymmetry and Nonperturbative beta Functions}},  {\em Phys. Lett. B} {\bf 206} (1988) 75--80.

\bibitem{Seiberg:1993vc}
N.~Seiberg, {\it {Naturalness versus supersymmetric nonrenormalization theorems}},  {\em Phys. Lett. B} {\bf 318} (1993) 469--475, [\href{http://arxiv.org/abs/hep-ph/9309335}{{\tt hep-ph/9309335}}].

\bibitem{Kutasov:1995ve}
D.~Kutasov, {\it {A Comment on duality in N=1 supersymmetric nonAbelian gauge theories}},  {\em Phys. Lett. B} {\bf 351} (1995) 230--234, [\href{http://arxiv.org/abs/hep-th/9503086}{{\tt hep-th/9503086}}].

\bibitem{Kutasov:1995np}
D.~Kutasov and A.~Schwimmer, {\it {On duality in supersymmetric Yang-Mills theory}},  {\em Phys. Lett. B} {\bf 354} (1995) 315--321, [\href{http://arxiv.org/abs/hep-th/9505004}{{\tt hep-th/9505004}}].

\bibitem{Kutasov:1995ss}
D.~Kutasov, A.~Schwimmer, and N.~Seiberg, {\it {Chiral rings, singularity theory and electric - magnetic duality}},  {\em Nucl. Phys. B} {\bf 459} (1996) 455--496, [\href{http://arxiv.org/abs/hep-th/9510222}{{\tt hep-th/9510222}}].

\bibitem{Lee:1998bxa}
S.~Lee, S.~Minwalla, M.~Rangamani, and N.~Seiberg, {\it {Three point functions of chiral operators in D = 4, N=4 SYM at large N}},  {\em Adv. Theor. Math. Phys.} {\bf 2} (1998) 697--718, [\href{http://arxiv.org/abs/hep-th/9806074}{{\tt hep-th/9806074}}].

\bibitem{Aharony:1999ti}
O.~Aharony, S.~S. Gubser, J.~M. Maldacena, H.~Ooguri, and Y.~Oz, {\it {Large N field theories, string theory and gravity}},  {\em Phys. Rept.} {\bf 323} (2000) 183--386, [\href{http://arxiv.org/abs/hep-th/9905111}{{\tt hep-th/9905111}}].

\bibitem{Romelsberger:2007ec}
C.~Romelsberger, {\it {Calculating the Superconformal Index and Seiberg Duality}},  \href{http://arxiv.org/abs/0707.3702}{{\tt arXiv:0707.3702}}.

\bibitem{Dolan:2008qi}
F.~A. Dolan and H.~Osborn, {\it {Applications of the Superconformal Index for Protected Operators and q-Hypergeometric Identities to N=1 Dual Theories}},  {\em Nucl. Phys. B} {\bf 818} (2009) 137--178, [\href{http://arxiv.org/abs/0801.4947}{{\tt arXiv:0801.4947}}].

\bibitem{Eager:2018dsx}
R.~Eager, I.~Saberi, and J.~Walcher, {\it {Nilpotence varieties}},  {\em Annales Henri Poincare} {\bf 22} (2021), no.~4 1319--1376, [\href{http://arxiv.org/abs/1807.03766}{{\tt arXiv:1807.03766}}].

\bibitem{Elliott:2020ecf}
C.~Elliott, P.~Safronov, and B.~R. Williams, {\it {A taxonomy of twists of supersymmetric Yang{\textendash}Mills theory}},  {\em Selecta Math.} {\bf 28} (2022), no.~4 73, [\href{http://arxiv.org/abs/2002.10517}{{\tt arXiv:2002.10517}}].

\bibitem{Budzik:2023xbr}
K.~Budzik, D.~Gaiotto, J.~Kulp, B.~R. Williams, J.~Wu, and M.~Yu, {\it {Semi-chiral operators in 4d $ \mathcal{N} $ = 1 gauge theories}},  {\em JHEP} {\bf 05} (2024) 245, [\href{http://arxiv.org/abs/2306.01039}{{\tt arXiv:2306.01039}}].

\bibitem{Gaiotto:2024gii}
D.~Gaiotto, J.~Kulp, and J.~Wu, {\it {Higher operations in perturbation theory}},  {\em JHEP} {\bf 05} (2025) 230, [\href{http://arxiv.org/abs/2403.13049}{{\tt arXiv:2403.13049}}].

\bibitem{Budzik:2022mpd}
K.~Budzik, D.~Gaiotto, J.~Kulp, J.~Wu, and M.~Yu, {\it {Feynman diagrams in four-dimensional holomorphic theories and the Operatope}},  {\em JHEP} {\bf 07} (2023) 127, [\href{http://arxiv.org/abs/2207.14321}{{\tt arXiv:2207.14321}}].

\bibitem{Bomans:2025klo}
P.~Bomans, N.~Garner, B.~R. Williams, and J.~Wu, {\it {Unravelling the Holomorphic Twist II: Anomalies and Extended Supersymmetry}},  \href{http://arxiv.org/abs/2509.16737}{{\tt arXiv:2509.16737}}.

\bibitem{Konishi:1983hf}
K.~Konishi, {\it {Anomalous Supersymmetry Transformation of Some Composite Operators in SQCD}},  {\em Phys. Lett. B} {\bf 135} (1984) 439--444.

\bibitem{Konishi:1985tu}
K.-i. Konishi and K.-i. Shizuya, {\it {Functional Integral Approach to Chiral Anomalies in Supersymmetric Gauge Theories}},  {\em Nuovo Cim. A} {\bf 90} (1985) 111.

\bibitem{Gui:2025dqp}
Z.~Gui, M.~Wang, and B.~R. Williams, {\it {Higher-dimensional Chiral Algebras in the Jouanolou Model}},  \href{http://arxiv.org/abs/2510.26608}{{\tt arXiv:2510.26608}}.

\bibitem{Costello:2013zra}
K.~Costello, {\it {Supersymmetric gauge theory and the Yangian}},  \href{http://arxiv.org/abs/1303.2632}{{\tt arXiv:1303.2632}}.

\bibitem{Saberi:2019ghy}
I.~Saberi and B.~R. Williams, {\it {Twisted characters and holomorphic symmetries}},  {\em Lett. Math. Phys.} {\bf 110} (2020), no.~10 2779--2853, [\href{http://arxiv.org/abs/1906.04221}{{\tt arXiv:1906.04221}}].

\bibitem{faonte2019higher}
G.~Faonte, B.~Hennion, and M.~Kapranov, {\it Higher kac--moody algebras and moduli spaces of g-bundles},  {\em Advances in Mathematics} {\bf 346} (2019) 389--466, [\href{http://arxiv.org/abs/1701.01368}{{\tt arXiv:1701.01368}}].

\bibitem{Gwilliam:2018lpo}
O.~Gwilliam and B.~R. Williams, {\it {Higher Kac{\textendash}Moody algebras and symmetries of holomorphic field theories}},  {\em Adv. Theor. Math. Phys.} {\bf 25} (2021), no.~1 129--239, [\href{http://arxiv.org/abs/1810.06534}{{\tt arXiv:1810.06534}}].

\bibitem{Saberi:2019fkq}
I.~Saberi and B.~R. Williams, {\it {Superconformal Algebras and Holomorphic Field Theories}},  {\em Annales Henri Poincare} {\bf 24} (2023), no.~2 541--604, [\href{http://arxiv.org/abs/1910.04120}{{\tt arXiv:1910.04120}}].

\bibitem{Bomans:2023mkd}
P.~Bomans and J.~Wu, {\it {Unravelling the Holomorphic Twist: Central Charges}},  {\em Commun. Math. Phys.} {\bf 405} (2024), no.~12 290, [\href{http://arxiv.org/abs/2311.04304}{{\tt arXiv:2311.04304}}].

\bibitem{Williams:2024mcc}
B.~R. Williams, {\it {The local cohomology of vector fields}},  \href{http://arxiv.org/abs/2405.05174}{{\tt arXiv:2405.05174}}.

\bibitem{Scheinpflug:2024mtn}
J.~Scheinpflug, {\it {The semi-chiral ring of supersymmetric $\phi^4$ theory as a representation}},  \href{http://arxiv.org/abs/2410.12980}{{\tt arXiv:2410.12980}}.

\bibitem{Chen:2025ujx}
H.~Chen and J.~Liniado, {\it {Infinite Dimensional Topological-Holomorphic Symmetry in Three-Dimensions}},  \href{http://arxiv.org/abs/2507.01858}{{\tt arXiv:2507.01858}}.

\bibitem{Loday1984CyclicHA}
J.-L. Loday and D.~G. Quillen, {\it Cyclic homology and the lie algebra homology of matrices},  {\em Commentarii Mathematici Helvetici} {\bf 59} (1984) 565--591.

\bibitem{Tsygan1983TheHO}
B.~L. Tsygan, {\it The homology of matrix lie algebras over rings and the hochschild homology},  {\em Russian Mathematical Surveys} {\bf 38} (1983) 198 -- 199.

\bibitem{Chang:2022mjp}
C.-M. Chang and Y.-H. Lin, {\it {Words to describe a black hole}},  {\em JHEP} {\bf 02} (2023) 109, [\href{http://arxiv.org/abs/2209.06728}{{\tt arXiv:2209.06728}}].

\bibitem{Choi:2022caq}
S.~Choi, S.~Kim, E.~Lee, and J.~Park, {\it {The shape of non-graviton operators for SU(2)}},  {\em JHEP} {\bf 09} (2024) 029, [\href{http://arxiv.org/abs/2209.12696}{{\tt arXiv:2209.12696}}].

\bibitem{Choi:2023znd}
S.~Choi, S.~Kim, E.~Lee, S.~Lee, and J.~Park, {\it {Towards quantum black hole microstates}},  {\em JHEP} {\bf 11} (2023) 175, [\href{http://arxiv.org/abs/2304.10155}{{\tt arXiv:2304.10155}}]. [Erratum: JHEP 03, 091 (2025)].

\bibitem{Budzik:2023vtr}
K.~Budzik, H.~Murali, and P.~Vieira, {\it {Following Black Hole States}},  \href{http://arxiv.org/abs/2306.04693}{{\tt arXiv:2306.04693}}.

\bibitem{Choi:2023vdm}
J.~Choi, S.~Choi, S.~Kim, J.~Lee, and S.~Lee, {\it {Finite N black hole cohomologies}},  {\em JHEP} {\bf 12} (2024) 029, [\href{http://arxiv.org/abs/2312.16443}{{\tt arXiv:2312.16443}}].

\bibitem{Chang:2024zqi}
C.-M. Chang and Y.-H. Lin, {\it {Holographic covering and the fortuity of black holes}},  \href{http://arxiv.org/abs/2402.10129}{{\tt arXiv:2402.10129}}.

\bibitem{deMelloKoch:2024pcs}
R.~de~Mello~Koch, M.~Kim, S.~Kim, J.~Lee, and S.~Lee, {\it {Brane-fused black hole operators}},  {\em JHEP} {\bf 07} (2025) 216, [\href{http://arxiv.org/abs/2412.08695}{{\tt arXiv:2412.08695}}].

\bibitem{Gaikwad:2025ugk}
A.~Gaikwad, T.~Kibe, S.~van Leuven, and K.~Mathieson, {\it {To gauge or to double gauge? Matrix models, global symmetry, and black hole cohomologies}},  \href{http://arxiv.org/abs/2512.02103}{{\tt arXiv:2512.02103}}.

\bibitem{Chang:2024lxt}
C.-M. Chang, Y.~Chen, B.~S. Sia, and Z.~Yang, {\it {Fortuity in SYK models}},  {\em JHEP} {\bf 08} (2025) 003, [\href{http://arxiv.org/abs/2412.06902}{{\tt arXiv:2412.06902}}].

\bibitem{Chen:2025sum}
Y.~Chen, {\it {Fortuity with a single matrix}},  \href{http://arxiv.org/abs/2511.00790}{{\tt arXiv:2511.00790}}.

\bibitem{Chang:2025rqy}
C.-M. Chang, Y.-H. Lin, and H.~Zhang, {\it {Fortuity in the D1-D5 system}},  \href{http://arxiv.org/abs/2501.05448}{{\tt arXiv:2501.05448}}.

\bibitem{Kim:2025vup}
S.~Kim, J.~Lee, S.~Lee, and H.~Oh, {\it {BPS phases and fortuity in higher spin holography}},  \href{http://arxiv.org/abs/2511.03105}{{\tt arXiv:2511.03105}}.

\bibitem{Chang:2025wgo}
C.-M. Chang and H.~Zhang, {\it {Fortuity and R-charge concentration in the D1-D5 CFT}},  \href{http://arxiv.org/abs/2511.23294}{{\tt arXiv:2511.23294}}.

\bibitem{belin2025fortuityabjm}
A.~Belin, P.~Singh, R.~Vadala, and A.~Zaffaroni, {\it {Fortuity in ABJM}},  \href{http://arxiv.org/abs/2512.04146}{{\tt arXiv:2512.04146}}.

\bibitem{Chang:2025mqp}
C.-M. Chang and Y.-H. Lin, {\it {Violation of S-duality in classical $Q$-cohomology}},  \href{http://arxiv.org/abs/2510.24008}{{\tt arXiv:2510.24008}}.

\bibitem{Choi:2025bhi}
J.~Choi and E.~Lee, {\it {Konishi lifts a black hole}},  \href{http://arxiv.org/abs/2511.09519}{{\tt arXiv:2511.09519}}.

\bibitem{Grant:2008sk}
L.~Grant, P.~A. Grassi, S.~Kim, and S.~Minwalla, {\it {Comments on 1/16 BPS Quantum States and Classical Configurations}},  {\em JHEP} {\bf 05} (2008) 049, [\href{http://arxiv.org/abs/0803.4183}{{\tt arXiv:0803.4183}}].

\bibitem{Gadde:2025yoa}
A.~Gadde, E.~Lee, R.~Raj, and S.~Tomar, {\it {Probing Non-Graviton Spectra in $\mathcal{N}=4$ SYM via BMN truncation and S-Duality}},  \href{http://arxiv.org/abs/2506.13887}{{\tt arXiv:2506.13887}}.

\bibitem{Chang:2013fba}
C.-M. Chang and X.~Yin, {\it {1/16 BPS states in $\mathcal N=$ 4 super-Yang-Mills theory}},  {\em Phys. Rev. D} {\bf 88} (2013), no.~10 106005, [\href{http://arxiv.org/abs/1305.6314}{{\tt arXiv:1305.6314}}].

\bibitem{Beem:2013sza}
C.~Beem, M.~Lemos, P.~Liendo, W.~Peelaers, L.~Rastelli, and B.~C. van Rees, {\it {Infinite Chiral Symmetry in Four Dimensions}},  {\em Commun. Math. Phys.} {\bf 336} (2015), no.~3 1359--1433, [\href{http://arxiv.org/abs/1312.5344}{{\tt arXiv:1312.5344}}].

\bibitem{Chang:2023ywj}
C.-M. Chang, Y.-H. Lin, and J.~Wu, {\it {On $\frac{1}{8}$-BPS black holes and the chiral algebra of $\mathcal{N}=4$ SYM}},  {\em Adv. Theor. Math. Phys.} {\bf 28} (2024), no.~7 2431--2489, [\href{http://arxiv.org/abs/2310.20086}{{\tt arXiv:2310.20086}}].

\bibitem{Axelrod:1991vq}
S.~Axelrod and I.~M. Singer, {\it {Chern-Simons perturbation theory}},  in {\em {International Conference on Differential Geometric Methods in Theoretical Physics}}, pp.~3--45, 1991.
\newblock \href{http://arxiv.org/abs/hep-th/9110056}{{\tt hep-th/9110056}}.

\bibitem{williams2020renormalization}
B.~R. Williams, {\it Renormalization for holomorphic field theories},  {\em Communications in Mathematical Physics} {\bf 374} (2020), no.~3 1693--1742, [\href{http://arxiv.org/abs/1809.02661}{{\tt arXiv:1809.02661}}].

\bibitem{Novikov:1983ee}
V.~A. Novikov, M.~A. Shifman, A.~I. Vainshtein, and V.~I. Zakharov, {\it {Instanton Effects in Supersymmetric Theories}},  {\em Nucl. Phys. B} {\bf 229} (1983) 407.

\bibitem{Hori:2003ic}
K.~Hori, S.~Katz, A.~Klemm, R.~Pandharipande, R.~Thomas, C.~Vafa, R.~Vakil, and E.~Zaslow, {\em {Mirror symmetry}}, vol.~1 of {\em Clay mathematics monographs}.
\newblock AMS, Providence, USA, 2003.

\bibitem{Witten:2003ye}
E.~Witten, {\it {Chiral ring of Sp(N) and SO(N) supersymmetric gauge theory in four-dimensions}},  \href{http://arxiv.org/abs/hep-th/0302194}{{\tt hep-th/0302194}}.

\bibitem{Etingof:2003dd}
P.~Etingof and V.~Kac, {\it {On the Cachazo-Douglas-Seiberg-Witten conjecture for simple Lie algebras}},  \href{http://arxiv.org/abs/math/0305175}{{\tt math/0305175}}.

\bibitem{Cederwall:2023lev}
M.~Cederwall and G.~Ferretti, {\it {The Chiral Ring of D=4, N=1 SYM with Exceptional Gauge Groups}},  {\em Fortsch. Phys.} {\bf 72} (2024), no.~5 2400027, [\href{http://arxiv.org/abs/2311.12119}{{\tt arXiv:2311.12119}}].

\bibitem{Dorey:2002ik}
N.~Dorey, T.~J. Hollowood, V.~V. Khoze, and M.~P. Mattis, {\it {The Calculus of many instantons}},  {\em Phys. Rept.} {\bf 371} (2002) 231--459, [\href{http://arxiv.org/abs/hep-th/0206063}{{\tt hep-th/0206063}}].

\bibitem{susyGTMM}
R.~Argurio, G.~Ferretti, and R.~Heise, {\it {An Introduction to supersymmetric gauge theories and matrix models}},  {\em Int. J. Mod. Phys. A} {\bf 19} (2004) 2015--2078, [\href{http://arxiv.org/abs/hep-th/0311066}{{\tt hep-th/0311066}}].

\bibitem{Seiberg:2002jq}
N.~Seiberg, {\it {Adding fundamental matter to `Chiral rings and anomalies in supersymmetric gauge theory'}},  {\em JHEP} {\bf 01} (2003) 061, [\href{http://arxiv.org/abs/hep-th/0212225}{{\tt hep-th/0212225}}].

\bibitem{Tachikawa:2018sae}
Y.~Tachikawa, {\it {Lectures on 4d N=1 dynamics and related topics}},  \href{http://arxiv.org/abs/1812.08946}{{\tt arXiv:1812.08946}}.

\bibitem{Dumitrescu:2011iu}
T.~T. Dumitrescu and N.~Seiberg, {\it {Supercurrents and Brane Currents in Diverse Dimensions}},  {\em JHEP} {\bf 07} (2011) 095, [\href{http://arxiv.org/abs/1106.0031}{{\tt arXiv:1106.0031}}].

\bibitem{Closset:2013vra}
C.~Closset, T.~T. Dumitrescu, G.~Festuccia, and Z.~Komargodski, {\it {The Geometry of Supersymmetric Partition Functions}},  {\em JHEP} {\bf 01} (2014) 124, [\href{http://arxiv.org/abs/1309.5876}{{\tt arXiv:1309.5876}}].

\bibitem{Closset:2014uda}
C.~Closset, T.~T. Dumitrescu, G.~Festuccia, and Z.~Komargodski, {\it {From Rigid Supersymmetry to Twisted Holomorphic Theories}},  {\em Phys. Rev. D} {\bf 90} (2014), no.~8 085006, [\href{http://arxiv.org/abs/1407.2598}{{\tt arXiv:1407.2598}}].

\bibitem{Dumitrescu:2016ltq}
T.~T. Dumitrescu, {\it {An introduction to supersymmetric field theories in curved space}},  {\em J. Phys. A} {\bf 50} (2016), no.~44 443005, [\href{http://arxiv.org/abs/1608.02957}{{\tt arXiv:1608.02957}}].

\bibitem{Yonekura:2010mc}
K.~Yonekura, {\it {Notes on Operator Equations of Supercurrent Multiplets and Anomaly Puzzle in Supersymmetric Field Theories}},  {\em JHEP} {\bf 09} (2010) 049, [\href{http://arxiv.org/abs/1004.1296}{{\tt arXiv:1004.1296}}].

\bibitem{Scheinpflug:2025sqn}
J.~Scheinpflug, M.~Schnabl, and J.~Vo{\v{s}}mera, {\it {Observables of boundary RG flows from string field theory}},  \href{http://arxiv.org/abs/2510.07155}{{\tt arXiv:2510.07155}}.

\bibitem{Witten:1992yj}
E.~Witten and B.~Zwiebach, {\it {Algebraic structures and differential geometry in 2-D string theory}},  {\em Nucl. Phys. B} {\bf 377} (1992) 55--112, [\href{http://arxiv.org/abs/hep-th/9201056}{{\tt hep-th/9201056}}].

\bibitem{Lian:1992mn}
B.~H. Lian and G.~J. Zuckerman, {\it {New perspectives on the BRST algebraic structure of string theory}},  {\em Commun. Math. Phys.} {\bf 154} (1993) 613--646, [\href{http://arxiv.org/abs/hep-th/9211072}{{\tt hep-th/9211072}}].

\bibitem{Getzler:1994yd}
E.~Getzler, {\it {Batalin-Vilkovisky algebras and two-dimensional topological field theories}},  {\em Commun. Math. Phys.} {\bf 159} (1994) 265--285, [\href{http://arxiv.org/abs/hep-th/9212043}{{\tt hep-th/9212043}}].

\bibitem{Beem:2018fng}
C.~Beem, D.~Ben-Zvi, M.~Bullimore, T.~Dimofte, and A.~Neitzke, {\it {Secondary products in supersymmetric field theory}},  {\em Annales Henri Poincare} {\bf 21} (2020), no.~4 1235--1310, [\href{http://arxiv.org/abs/1809.00009}{{\tt arXiv:1809.00009}}].

\bibitem{Oh:2019mcg}
J.~Oh and J.~Yagi, {\it {Poisson vertex algebras in supersymmetric field theories}},  {\em Lett. Math. Phys.} {\bf 110} (2020), no.~8 2245--2275, [\href{http://arxiv.org/abs/1908.05791}{{\tt arXiv:1908.05791}}].

\bibitem{Garner:2022its}
N.~Garner and N.~M. Paquette, {\it {Mathematics of String Dualities}},  {\em PoS} {\bf TASI2021} (2023) 007, [\href{http://arxiv.org/abs/2204.01914}{{\tt arXiv:2204.01914}}].

\bibitem{Gaiotto:2015aoa}
D.~Gaiotto, G.~W. Moore, and E.~Witten, {\it {Algebra of the Infrared: String Field Theoretic Structures in Massive ${\cal N}=(2,2)$ Field Theory In Two Dimensions}},  \href{http://arxiv.org/abs/1506.04087}{{\tt arXiv:1506.04087}}.

\bibitem{Kontsevich:1997vb}
M.~Kontsevich, {\it {Deformation quantization of Poisson manifolds. 1.}},  {\em Lett. Math. Phys.} {\bf 66} (2003) 157--216, [\href{http://arxiv.org/abs/q-alg/9709040}{{\tt q-alg/9709040}}].

\bibitem{Balduf:2024wwp}
P.-H. Balduf and D.~Gaiotto, {\it {Combinatorial proof of a non-renormalization theorem}},  {\em JHEP} {\bf 05} (2025) 120, [\href{http://arxiv.org/abs/2408.03192}{{\tt arXiv:2408.03192}}].

\bibitem{Wang:2024sqm}
M.~Wang, {\it {Feynman Graph Integrals on $\mathbb {C}^d$}},  {\em Commun. Math. Phys.} {\bf 406} (2025), no.~5 116, [\href{http://arxiv.org/abs/2401.08113}{{\tt arXiv:2401.08113}}].

\bibitem{Wang:2024tjf}
M.~Wang and B.~R. Williams, {\it {Factorization algebras from topological-holomorphic field theories}},  \href{http://arxiv.org/abs/2407.08667}{{\tt arXiv:2407.08667}}.

\bibitem{Wang:2025rmu}
M.~Wang and J.~Yan, {\it {Feynman Graph Integrals on K{\"a}hler Manifolds}},  \href{http://arxiv.org/abs/2507.09170}{{\tt arXiv:2507.09170}}.

\bibitem{henneberg1911graphische}
L.~Henneberg, {\em Die graphische Statik der starren Systeme}, vol.~31.
\newblock BG Teubner, 1911.

\bibitem{pollaczek1927gliederung}
H.~Pollaczek-Geiringer, {\it {\"U}ber die gliederung ebener fachwerke},  {\em ZAMM-Journal of Applied Mathematics and Mechanics/Zeitschrift f{\"u}r Angewandte Mathematik und Mechanik} {\bf 7} (1927), no.~1 58--72.

\bibitem{laman1970graphs}
G.~Laman, {\it On graphs and rigidity of plane skeletal structures},  {\em Journal of Engineering mathematics} {\bf 4} (1970), no.~4 331--340.

\bibitem{Johansen:1994aw}
A.~Johansen, {\it {Twisting of $N=1$ SUSY gauge theories and heterotic topological theories}},  {\em Int. J. Mod. Phys. A} {\bf 10} (1995) 4325--4358, [\href{http://arxiv.org/abs/hep-th/9403017}{{\tt hep-th/9403017}}].

\bibitem{Costello:2011np}
K.~J. Costello, {\it {Notes on supersymmetric and holomorphic field theories in dimensions 2 and 4}},  {\em Pure Appl. Math. Quart.} {\bf 09} (2013), no.~1 73--165, [\href{http://arxiv.org/abs/1111.4234}{{\tt arXiv:1111.4234}}].

\end{thebibliography}\endgroup

\end{document}